\documentclass[12pt,a4paper]{article}    
 \usepackage[skins,theorems]{tcolorbox}
\tcbset{highlight math style={enhanced,
  colframe=red,colback=white,arc=0pt,boxrule=1pt}}
 \usepackage[bookmarksopen, bookmarksnumbered, bookmarksopenlevel=2]{hyperref}
  \usepackage{tikz}
  \usepackage{tikz-3dplot}
 \usetikzlibrary{calc}
 \usetikzlibrary{decorations} %
 \usepackage[english]{babel}
 \usepackage[toc,page]{appendix}
 \usepackage{amsmath}
 \usepackage{amssymb}
 \usepackage{graphicx}
 \usepackage{hhline}
 \usepackage[bf]{caption}
\usepackage[vcentermath]{youngtab}
\usepackage{geometry}
\usepackage{slashed}
\usepackage{color}
\usepackage{stackrel}
\usepackage{tikz-cd} 
\usepackage{cancel} 
\usepackage{booktabs} 
\usepackage{sectsty}
\sectionfont{\fontsize{15}{22}\selectfont}
\subsectionfont{\fontsize{13}{15}\selectfont}
\usepackage{empheq}
\usepackage{arydshln}

\usepackage[numbers, square, comma,compress]{natbib} 

\newcommand{\bqa}{\begin{eqnarray}}
\newcommand{\eqa}{\end{eqnarray}}

\newcommand{\nn}{\nonumber}

\usepackage{tikz}
\usetikzlibrary{shadows} 
\usepackage[framemethod=tikz]{mdframed}
\global\mdfdefinestyle{myboxstyle}{%
  shadow=true,
  linecolor=black,
  shadowcolor=black,
  shadowsize=6pt,
  nobreak=false,
  innertopmargin=10pt,
  innerbottommargin=10pt,
  leftmargin=5pt,
  rightmargin=5pt,
  needspace=1cm,
  skipabove=10pt,
  skipbelow=15pt,
  middlelinewidth=1pt,
  afterlastframe={\vspace{5pt}},
  aftersingleframe={\vspace{5pt}},
  tikzsetting={%
draw=black,
very thick} }
\newmdenv[style=myboxstyle]{whitebox} \newmdenv[style=myboxstyle,backgroundcolor=black!20]{graybox}
\newmdenv[style=myboxstyle,nobreak=true]{blockwhitebox}
\newmdenv[style=myboxstyle,backgroundcolor=black!20,nobreak=true]{blockgraybox}
\newmdenv[nobreak=true,hidealllines=true]{blockbox}

\hypersetup{
    pdftitle={},
    pdfauthor={},
    pdfsubject={}
}
\numberwithin{equation}{section}
\numberwithin{table}{section}

\makeatletter

\newcommand{\be}{\begin{equation}}
\newcommand{\ee}{\end{equation}}
\newcommand{\beq}{\begin{equation}}
\newcommand{\eeq}{\end{equation}}
\newcommand{\ba}{\begin{aligned}}
\newcommand{\ea}{\end{aligned}}

\newcommand{\bea}{\begin{eqnarray}}
\newcommand{\eea}{\end{eqnarray}}

\newcommand{\cO}{\mathcal{O}}

\newcommand{\cF}{\mathcal{F}}

\newcommand{\cS}{\mathcal{S}}

\newcommand{\cM}{\mathcal M}

\newcommand\bi{\begin{itemize}}
\newcommand\ei{\end{itemize}}

\def\Im{\mathop{\mathrm{Im}}\nolimits}

\def\unit{{1\kern-.65ex {\rm l}}}
\def\1{{1\kern-.65ex {\rm l}}}

\def\IZ{\mathbb{Z}}
\def\IP{\mathbb{P}}
\def\IL{\mathbb{L}}

\def\hilb{{\rm Hilb}}
\def\sym{{\rm Sym}}
\def\al{{\alpha}}
\def\bet{{\beta}}

\def\pt{{\rm pt}}

\def\IH{\mathbb{H}}
\def\IE{\mathbb{E}}
\def\IC{\mathbb{C}}
\def\IQ{\mathbb{Q}}

\def\GW{{\rm GW}\!}
\def\GV{{\rm GV}}
\def\hG{\widehat{G}}
\def\tG{\widetilde{G}}
\def\htG{\widehat{\tG}{}}
\def\bark {{\bar k}}
\def\shalf{\relax{\textstyle {1 \over 2}}\displaystyle}

\newcount\hour \newcount\minute
\hour=\time \divide \hour by 60
\minute=\time
\def\now{%
\ifnum \hour<13
  \ifnum \hour=0 \advance \hour by 12 \number\hour:\else \number\hour:\fi%
     \ifnum \minute<10 0\fi%
     \number\minute%
\ A.M.%
\else \advance \hour by -12 \number\hour:%
  \ifnum \minute<10 0\fi%
  \number\minute%
  \ P.M.%
\fi%
}

\def\fnote#1#2{\begingroup\def\thefootnote{#1}\footnote{#2}
     \addtocounter{footnote}{-1}\endgroup}

\makeatother

\begin{document}

\begin{flushright}

{\tt\normalsize CERN-TH-2023-098}\\

\end{flushright}

\vskip 20 mm

\begin{center}
{\large \bf
Gromov-Witten/Hilbert versus AdS$_3$/CFT$_2$ Correspondence\\
} 
\vskip 30 mm

Wolfgang Lerche${}^{*}$

\vskip 7 mm
{\it CERN, Theory Department, \\ 1, Esplanade des Particules, Geneva 23, CH-1211, Switzerland} \\[3 mm]

\fnote{}{${}^{*}$wolfgang.lerche at cern.ch}

\end{center}

\vskip 15 mm
\begin{abstract}
\vskip 5 mm
We consider the boundary dual of AdS$_3\times S^3\times K3$
for NS5-flux $Q_5^{NS}=1$, which is described by a sigma model with target space given
by the $d$-fold symmetric product of $K3$.
Building on results in algebraic geometry,  we address the problem of deforming it
away from the orbifold point from the viewpoint of topological strings.
We propose how the 't~Hooft expansion can be geometrized in terms of 
Gromov-Witten invariants and, in favorable settings, 
how it can be summed up to all orders in closed form.
We consider an explicit example in detail for which we discuss the genus expansion
around the orbifold point, as well as the divergence in the strong coupling regime. 
We find that within the domain of convergence, scale separation does not occur.
However, in order for the mathematical framework to be applicable in the first place, 
we need to consider ``reduced'' Gromov-Witten invariants that fit, as we argue, naturally
to topologically twisted $N=4$ strings.  There are some caveats and thus
to what extent this toy model captures  the physics of strings on AdS$_3\times S^3\times K3$ 
remains to be seen. 
\end{abstract}

\vfill

\thispagestyle{empty}
\setcounter{page}{0}

\setcounter{page}{1}
\newpage 

\tableofcontents
\goodbreak

\section{Introduction} \label{sec_intro}

\subsection{Motivation} 

One of the best understood and tested instances of the AdS/CFT Correspondence \cite{Maldacena:1997re} is the 
near horizon limit of the F1NS5 (resp.~the dual D1D5) system. It corresponds to
a Type IIB string compactification on AdS$_3\times S^3\times M_4$, where $M_4=T^4$ or $K3$.
It was found that the system drastically simplifies\footnote{Most results were found for $M_4=T^4$ but one expects the theory to behave analogously for
$M_4=K3$.}
 for one unit of NS5 flux, ie., $Q_5^{NS}=1$
\cite{Giribet:2018ada,Gaberdiel:2018rqv,Eberhardt:2018ouy,Eberhardt:2019ywk,Gaberdiel:2020ycd,Eberhardt:2020bgq,Bhat:2021dez,Gaberdiel:2022als}.
 The continuous spectrum of long strings collapses to the bottom, and strings nestled against the
boundary $\Sigma=\partial$AdS$_3$ become tensionless. Also, the world-sheet theory becomes
free and is to a large extent solvable. On the other hand, the boundary theory is 
described by a two dimensional sigma model with target space $\sym^N\!(M_4)$, where $N=Q_1^{NS}Q_5^{NS}$ for $M_4=T^4$ and $N=Q_1^{NS}Q_5^{NS}+1$ for $M_4=K3$ \cite{Larsen:1999uk,Beccaria:2014qea}. 
Numerous tests, mainly for $M_4=T^4$,  confirmed the equality of correlation functions \cite{Jevicki:1998bm,Lunin:2000yv,Lunin:2001pw,Rastelli:2005xaa,Gaberdiel:2007vu,Dabholkar:2007ey,Taylor:2007hs,Giribet:2007wp,Pakman:2007hn,deBoer:2008ss,Pakman:2009zz,Pakman:2009ab,Cardona:2010qf,Kirsch:2011na,Baggio:2012rr,Rastelli:2019gtj,Dei:2019,Eberhardt:2020akk,Hikida:2020kil,Dei:2020zui,Knighton:2020kuh,Eberhardt:2021jvj,Lima:2021wrz,Dei:2021xgh,Dei:2021yom,AlvesLima:2022elo,Dei:2022pkr,Knighton:2022ipy,Gaberdiel:2022oeu,Iguri:2022pbp,Ashok:2023mow,Iguri:2023khc,Ashok:2023kkd} in the large $N$ limit.

Most of the work done so far deals with computations in the  tensionless regime which
is characterized by the singular orbifold $\sym^N\!(M_4)$. There is a universal marginal
operator, given by the twist field $\sigma_2$, 
which is associated with the blow-up mode of the orbifold singularity. One can perturb the theory by\footnote{For now, $u$ is a real parameter. Later  we will complexify it by adding a theta-angle,
such that $u=0$  corresponds to $\theta/2\pi=1/2$.  Note that $u=0$ then corresponds precisely to the theory with $Q_5^{NS}=1$, with no further shifts of background moduli.}
\be
 \delta S\ \sim\   u\int G_{-1}^-\bar G_{-1}^- \sigma_2\,,
 \ee
which desingularizes  $\sym^N\!(M_4)$ by deforming it into the Hilbert scheme of $N$ points on $M_4$,
$\hilb^N\!(M_4)$ (for an extensive review of the role of Hilbert schemes in the context of D-brane physics and black holes, see \cite{Dijkgraaf:1998gf}; for general aspects of this deformation, in particular from the supergravity point of view, see eg.~\cite{Giveon:1998ns,Berkovits:1999im,Larsen:1999uk,David:2002wn,OhlssonSax:2018hgc,Eberhardt:2018vho,Martinec:2022okx}).  
Importantly, it has been known since long that $u$ has characteristic features of a string coupling,
and in particular that it should lead to a genus expansion \cite{Dijkgraaf:1998xr}.
It cannot however coincide with the string coupling proper, as there is
a non-trivial genus expansion already at $u=0$. One of our goals is to clarify the role of $u$, and 
address its all-order expansion globally over the moduli space.

This problem is interesting from several perspectives, and was the initial motivation for this work.
 For example,  the extrapolation from $u=0$ to $u\rightarrow\infty$ touches upon
the question of scale separation \cite{Lust:2019zwm,Perlmutter:2020buo,Baume:2020dqd,Shiu:2022oti}.
The AdS and string scales are tied together by the supergravity formula \cite{Maldacena:1998bw,Berkovits:1999im,David:2002wn,OhlssonSax:2018hgc}:
 \be\label{Rads}
 \frac {R^2_{{\rm AdS}}}{\alpha'} \ \sim\ Q_5^{NS}{\sqrt{1+(g_{10} c_2)^2}}\,,
\ee
where $g_{10}=e^{\Phi}$ is the ten dimensional string coupling and
 $C_2^+ =: c_2(e^6\wedge e^7+e^8\wedge e^9)$ is a  self-dual RR two-form.
Assuming validity of the formula in this regime,
it implies that the scales are of the same order at the orbifold point where $Q_5^{NS}=1$
and $c_2=0$. Switching on $u$ amounts to switching on 
$c_2$ \cite{Larsen:1999uk,David:2002wn,OhlssonSax:2018hgc}, 
 so that the size of AdS would increase as $ \frac {R^2_{{\rm AdS}}}{\alpha'} \propto u$ for large $u$,
  but can one make it larger than the size of $M_4$ in order to reach the supergravity regime?
 
Relatedly one could ask how the deformation from $u=0$ to $u\rightarrow\infty$
fits to the general picture of infinite 
distance limits \cite{Ooguri_2007} within the Swampland Program
 (see eg.~\cite{Palti:2019pca,vanBeest:2021lhn,Agmon:2022thq}) for reviews).
It is well established that near $u=0$ there is an infinite tower of weakly coupled tensionless strings
\cite{Sundborg:2000wp,Ferreira:2017pgt,Gaberdiel:2014cha,Gaberdiel:2015wpo,Baggio:2015jxa,Gaberdiel:2018rqv}, while
the general expectation is that at large $u$,
the theory should enter a strongly coupled supergravity regime \cite{Seiberg:1999xz}. 
However, since geometrically $u$ corresponds to the blow-up mode of a $\IZ_2$ orbifold singularity, the point $u=0$ 
corresponds to a finite distance singularity\footnote{Due to supersymmetry, singularites due to massless states ought to appear only in corrections to higher derivative terms.} at the center of the moduli space.
Thus there seems a bit of  tension
with the lore that  it is not at small but at large distance limits 
where  asymptotically massless towers of higher spin string or KK states should appear that are weakly coupled to gravity  (in some suitable duality frame)
\cite{Lust:2019zwm,Perlmutter:2020buo,Baume:2020dqd,Etheredge:2022opl,Baume:2023msm}.
A  viewpoint to reconcile this could be that the tensionless strings
at $u=0$ are in this respect analogous to the  
non-gravitational tensionless strings that arise at ADE singularities in the interior of the moduli space
of string compactifications; indeed the geometry at $u=0$ is strongly curved from the viewpoint of supergravity, and is not a weakly coupled gravitational theory {\it despite} having a perturbative genus expansion in an appropriate duality frame.

Clearly, in order to address such kind of questions one would like to have analytical control over 
the whole of the $u$-moduli space, including the regime where $u$ is large.
There have been numerous computations in conformal perturbation theory, see for example refs.~\cite{Avery:2010er,Avery:2010hs,Burrington:2012yq,Carson:2014yxa,Carson:2014ena,Gaberdiel:2015uca,Carson:2015ohj,
Carson:2016uwf,Burrington:2017jhh,Keller:2019yrr,Bufalini:2022toj,Guo:2019ady,Lima:2020boh,   Lima:2020nnx,Guo:2020gxm,Lima:2020kek,Benjamin:2021zkn,Eberhardt:2021vsx,Apolo:2022fya,Guo:2022ifr,Guo:2022zpn,
Fiset:2022erp,Eberhardt:2018exh}.  
However, standard CFT methods are typically limited to the regime near $u=0$, and the situation is aggravated by the complicated combinatorics of twist fields, which makes an all-genus expansion a tall order.

On the other hand, all-order expansions are routine in geometrical setups, such as moduli spaces of topological strings
on Calabi-Yau manifolds. 
Mirror symmetry and/or localization techniques allow (in BPS protected sectors) to sum up infinite orders in conformal perturbation theory in closed form, including all contact terms. 
Moreover, a geometrical modulus can turn into a string coupling constant in
another duality frame, which opens the door to all-genus computations.
 This is exactly what we propose to be the case here as well, namely that the geometrical
blow-up modulus $u$ on the CFT side will map to a perturbative 't~Hooft coupling 
$\lambda$ on the AdS side.

Certainly we expect such a geometric approach, if at all, 
 to capture the physics of the geometry
AdS$_3\times S^3\times M_4$ at best for $Q_5^{NS}=1$, where there is no continuum of long strings and the theory drastically simplifies \cite{Giribet:2018ada,Gaberdiel:2018rqv,Eberhardt:2018ouy,Eberhardt:2019ywk,Gaberdiel:2020ycd,Eberhardt:2020bgq,Bhat:2021dez,Gaberdiel:2022als}.
Indeed the theory has characteristic features of a topological $N=4$ string \cite{Dei:2020zui,Bhat:2021dez,Gaberdiel:2022als,Knighton:2022ipy,McStay:2023thk}, for which the sectors
AdS$_3\times S^3$ and $K3$  have separately vanishing central charges. In the hybrid formulation, the first factor
can be described by a ${\mathfrak{psu}}(1,1|2)_1$  WZW model with extra ghost fields,
while the latter is described by a topologically twisted $N=4$ sigma model \cite{Berkovits:1994vy,Berkovits:1999im} on $K3$. 
It is not clear to us to what degree this describes the full theory, but we expect it to capture 
at least a protected BPS subsector of it \cite{Dabholkar:2007ey,Pakman:2007hn, deBoer:2008ss, Baggio:2012rr, Gaberdiel:2022oeu, Martinec:2022okx,Bufalini:2022toj,Ashok:2023mow,Iguri:2023khc,Ashok:2023kkd},
thereby realizing what may be loosely called topological AdS/CFT Correspondence.\footnote{Note that several notions of a  topological AdS/CFT Correspondence exist, see eg.~\cite{Sugawara:1999fq,Rastelli:2005xaa,Costello:2020jbh,Li:2020zwo,Li:2020nei,Ashok:2023mow}.}

In the next section we will outline how such a topological AdS/CFT Correspondence could work by making use of some mathematical results for $M_4=K3$.

\subsection{AdS$_3$/CFT$_2$ and GW/Hilb Correspondences} \label{sec_GWHilb}

\begin{figure}[t!]
\centering
\includegraphics[width=15cm]{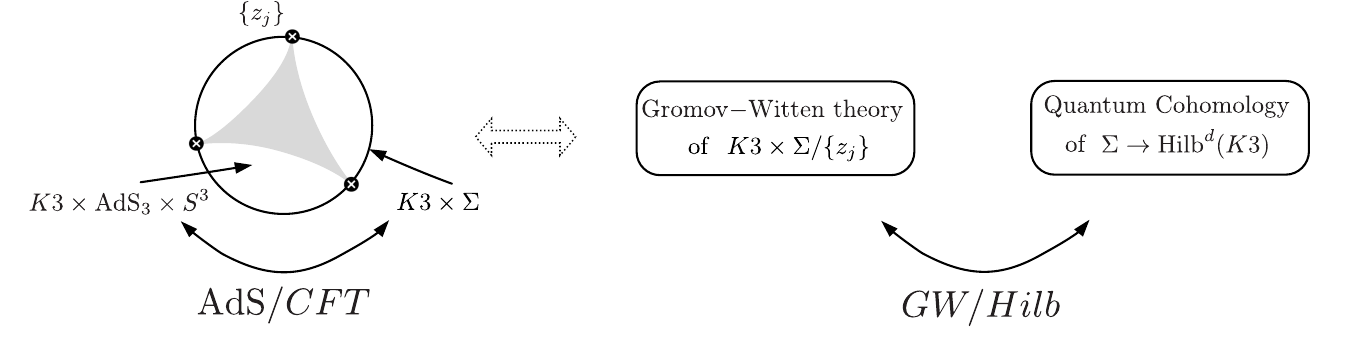}
\caption{Connecting AdS/CFT and GW/Hilb Correspondences, where $\Sigma=\partial {\rm AdS}_3$.
}
 \label{fig_correspondences}
\end{figure}

The mathematical key  statement on which we will base our discussion is a correspondence
between the quantum cohomology of the Hilbert scheme of $d$ points on $K3$,\footnote{Analogous correspondences exist also for geometries other than $K3$, see \cite{Oberdieck:2021beb} for a list of references.}  
 and 
the Gromov-Witten theory on $K3\times \Sigma$, which originates from the works of Oberdieck and Pandharipande \mbox{\cite{Ober_1406.1139,Ober_1605.05238,Oberdieck:2014aga,Oberdieck:2022khj}}.
Our purpose is to see whether this correspondence can offer new
insights into AdS$_3$/CFT$_2$ Correspondence in terms of algebraic geometry.
A schematic representation is shown in Figure~\ref{fig_correspondences}.

One is tempted to associate the ``Hilbert'' box to the far right of Figure~\ref{fig_correspondences} with
the boundary CFT, and relate the ``Gromov-Witten'' box via the dotted arrow to the topological
string theory at the boundary. 
The curve $\Sigma$ would thus play a dual role, namely as the world-sheet
of the 2d CFT sigma-model on $\hilb^d(K3)$, as well as (part of) the target space on the gravity side, namely as boundary of anti-de-Sitter space: $\Sigma=\partial {\rm AdS}_3$;
 we will focus on $\Sigma=\IP^1$ here.
The string world-sheets on the AdS side would then arise via branched coverings of degree $d$ of $\Sigma$,
so that the sum over branched coverings yields a perturbative genus expansion,
precisely in the spirit of the literature that we cited in the previous section. 

Note that the mathematical GW/Hilb Correspondence is an exact duality and holds for any $d\geq1$,  whereas the physical AdS$_3$/CFT$_2$ Correspondence, which relates the boundary
theory based on $K3\times \IP^1$ to the bulk of AdS$_3$,  holds primarily
in the large $N$ limit. An immediate question is whether these two correspondences could fit together
at their intersection $K3\times \IP^1$, 
 and what concrete benefits this connection could provide. Clearly, one
 new perspective the geometric approach 
could offer is that the counting of covering maps
is phrased in the language of GW invariants, thereby providing a
link to topological gravity and Hurwitz theory, which in turn has the potential for solving the genus expansion to all orders. 

In order to address these questions, there are several but related issues to be sorted out. 
One issue deals with the enumerative counting of holomorphic covering maps from the world-sheet
to $\IP^1$, another with the counting of holomorphic curves in $K3$ (an analogous
problem exists for $\hilb^d(K3)$ if we talk about the ``Hilbert'' side of the correspondence).
Their common problem is that Gromov-Witten theory is naturally based on holomorphic geometry while
the $N=(4,4)$ superconformal AdS background is based on hyperk\"ahler geometry
and so canonically has a harmonic rather than a holomorphic structure. Thus, connecting
these two correspondences requires at least 
projecting to a suitable holomorphic structure embedded in the
quaternionic hyperk\"ahler geometry, or in physical terms,  singling out a suitable
 $N=(2,2)$ superconformal  subsector of
the topologically twisted $N=(4,4)$ string on the boundary of AdS$_3\times S^3\times K3$.

Let us first consider $\IP^1=\partial$AdS$_3$, which is a boundary in
the full geometry 
so there is an interior into which world-sheets can in principle collapse.  
Thus an enumerative counting of holomorphic maps can make
 only sense if the string world-sheets are tied to the boundary so that there is a meaningful notion of them ``wrapping'' it. In fact it has been argued \cite{deBoer:1998gyt,Maldacena:2000hw,Eberhardt:2019ywk,Bhat:2021dez,Knighton:2022ipy} that world-sheets of long strings are pinned to the boundary at the branch points $\{z_j\}$,  but otherwise, especially for large twists, may contract to the interior; see Figure~\ref{fig_correspondences} for a sketchy visualization.
One may expect that in the present framework where one counts holomorphic coverings
of $\IP^1$ of degree $d$, all what matters is the topology of the branch points at the boundary.

There is indeed strong evidence that for $Q_5^{NS}=1$, the
world-sheets effectively do localize at the boundary via holomorphic maps. 
This mechanism has been recently understood \cite{Dei:2020zui,Bhat:2021dez,Gaberdiel:2022als,Knighton:2022ipy} 
by reformulating the ${\mathfrak{psu}}(1,1|2)_1$  WZW model that underlies the $N=(4,4)$ superconformal
quantum geometry of AdS$_3\times S^3$ in terms of
free twistor variables, and showing that the associated holomorphic twistorial maps localize at the boundary.
Moreover very recently it has been argued \cite{McStay:2023thk} that precisely for $Q_5^{NS}=1$
there appears a radial global symmetry which more or less
guarantees this to be true. One thus expects that the theory on AdS$_3\times S^3$ reduces 
to some version of topological strings on $\IP^1$ (whose precise nature needs to be clarified),
and this opens the possibility to count holomorphic covering maps 
via the Gromov-Witten/Hurwitz Correspondence \cite{OkouPandha0204305}, thereby producing a 
non-trivial genus expansion despite $\IP^1=\partial$AdS$_3$ being a boundary.

The other issue is that standard Gromov-Witten invariants of $K3$ (and of its associated Hilbert schemes) 
vanish identically, as a consequence of its hyperk\"ahler structure.
Indeed for a generic choice of complex structure,
there are no holomorphic curves of pure Hodge type (1,1) at all that could be counted.
In the mathematical literature one  defines instead {\it {reduced}} Gromov-Witten invariants \cite{Bryan1997TheEG, Maul_0705.1653,Maulik:2010jw,Klemm_0807.2477,Pandharipande:2014ooa}, for example
 by considering twistor families of complex structures on $K3$.
 For each such twistor family there exists a fixed
choice of complex structure for which curves become holomorphic so that
they can be counted in terms of non-vanishing invariants \cite{Bryan1997TheEG}.

Thus for both issues there is a common underlying structure that provides
the desired connection between the (not naturally holomorphic) hyperk\"ahler geometry and the counting of holomorphic maps from the world-sheet to $\partial$AdS$_3\times K3$.
We will propose later in Section~\ref{sec_topstring} 
how to translate the mathematical concept of reduced Gromov-Witten 
invariants to the language of topologically twisted $N=4$ strings \cite{Berkovits:1994vy,Berkovits:1999im}. 

While these reduced invariants count certain BPS states, the precise relationship between the 
correlators of the topological toy model and those of the physical theory on the AdS space remains unclear. 
This also touches upon the
compatibility with non-renormalization theorems \cite{deBoer:2008ss,Baggio:2012rr} 
that are known for extremal correlators~\cite{DHoker:1999jke}.
While we will make some comments in this respect in Sections~\ref{sec_topstring} and \ref{sec_summary},
we leave a more detailed analysis to future work. For now we stick in this paper to the right half of Figure~\ref{fig_correspondences}
and simply discuss generating functions of certain BPS invariants, without caring about their interpretation 
in terms of space-time correlators on the AdS side. Rather our focus will be
on the moduli space of the blow-up mode of the orbifold singularity and its genus expansion. 

In the next section we will collect a few underlying
mathematical concepts on which we will base our discussion. 
We won't add anything new here, but will in turn interpret some
of these  from a physical perspective.  The informed reader is invited to jump to
the key  Section~\ref{sec_GWH} which is devoted to a discussion of the Gromov-Witten/Hilbert
Correspondence. Subsequently, 
in Section \ref{sec_tHooft} we will  give an interpretation of the genus expansion in terms of a `t~Hooft expansion on top of the large-$N$ expansion.  In Section~\ref{sec_example} we discuss a simple
toy model and discuss in some detail the nature of the genus expansion, as well comment
on the strong coupling regime. Finally in Section~\ref{sec_summary} we recapitulate our
discussion and mention some loopholes of our proposal.

\vfil\eject

\section{Background material}\label{subsec_recap}

We recall some facts that are well-known in the mathematical literature, with focus on
the cohomology of the Hilbert scheme of $d$ points on $K3$,  $\hilb^d(K3)$,
and on what are called reduced Gromov-Witten invariants. Moreover, in order
to make this accessible from a physicist's perspective, we propose an interpretation
of the latter in terms of $N=4$ topological strings.
For details on the relevant algebraic geometry we refer eg.~to~
\cite{Pandharipande:2011jz,Ober_1406.1139,Ober_1605.05238,Oberdieck:2014aga,dedieu_02914652}. 

\subsection{(Co)homology of Hilbert Schemes} \label{sec_hilb}

The Hilbert scheme, $\hilb^d(K3)$, can be viewed as a deformation of the orbifold symmetric product
$\sym^d(K3)={K3}^{\otimes d}/S_d$  (see eg.,~\cite{Dijkgraaf:1998gf,Dijkgraaf:1999zd}).  It is a hyperk\"ahler manifold of complex dimension~$2d$. For $d>1$ its
second cohomology is well-known to be of dimension 23, which is one larger than that of a single $K3$:
\be
H^2(\hilb^d(K3),\IQ) = H^2(K3,\IQ) \oplus \IC\,[\gamma^{(\pi_A)}]\,,\ \ \ d>1\,.
\ee
The extra piece corresponds to the class of the blow-up mode of the $\IZ_2$ orbifold singularity on the small diagonal
in $\sym^d(K3)$ where two points collide. An explicit representative will be
given below in eq.~(\ref{sigma2}).  It has been recognized since long that the corresponding modulus,
$u$, plays the role of a string coupling constant (see eg., ref.~\cite{Dijkgraaf:1998xr}). This will be our main theme to which we will return later. 

The full (co-)homology of $\hilb^d(K3)$ can be given a convenient Fock space representation. 
We will be brief, details can be found eg.~in ref.~\cite{Ober_1605.05238}. The basic ingredients are the Nakajima creation operators
\be\label{nakaj}
p_{-a}(\al): \ H^*(\hilb^b(K3),\IQ)\ \longrightarrow H^*(\hilb^{a+b}(K3),\IQ)\,,\qquad   \al\in  H^*(K3)\,,
\ee
which lift cohomology elements  $\gamma$ to $p_{-a}(\al)\gamma$
while adding cycles of length $a$ to $\al$. In physical terms, the
$p_{-a}(\_)$ correspond to twist fields $\sigma_a$ that will
create ``long'' strings of length $a$ on the AdS side. In this way, one can systematically generate 
the full cohomology of $\hilb^d(K3)$ by repeatedly acting with $p_{-a}(\al)$ on the Fock vacuum, $\bf 1$. 
An arbitrary cohomology representative can be assembled by pairing
some (non-ordered) partition $\mu=[\mu_1,...\mu_\ell]$ of  size 
$d=\sum \mu_i$ and length $\ell(\mu) = d-\sum (\mu_i-1)$ 
with cohomology elements  $\alpha_i\in H^{q_i,q_i}(K3,\IQ)$. 
Altogether this form what is called a cohomology weighted partition:
\be
\pi\ =\ [(\mu_i,\al_i)]\,,\qquad i=1,\dots,\ell(\mu)\,.
\ee
Specifically:
\be\label{gammainsert}
\gamma^{(\pi )}\ =\ \mathfrak{z}(\mu)^{-1} \prod_i p_{-\mu_i}(\al_i) {\bf 1}\ \in\  H^{q ,q }(\hilb^d(K3),\IQ)  \,, 
\ee
with symmetry factor
\be\label{mathfreak}
\mathfrak{z}(\pi)\ =\ |{\rm Aut}(\pi)|\prod \mu_i
\ee 
that we don't write explicitly in the following.
As indicated, one can associate with $\gamma^{(\pi )}$  a total degree, or $R$-charge given by
\be\label{degree}
q \ =\ \sum_{i=1}^{\ell(\pi)}  \left(q_i+(\mu_i-1)\right)\,.
\ee
Alternatively, one can consider the Poincar\'e duals of (\ref{gammainsert})
which we use interchangeably with the same notation (except with subscripts) when it fits better.

The notation should become more clear when considering examples of
cohomology/curve classes that will be relevant for us. The first originates from a generic curve class $\bet\in H_2(K3,\IZ)$,
which will eventually be linked to a  K\"ahler modulus.  Its image in the Hilbert scheme
\be
C(\beta)\ \equiv \gamma_{(\pi_\bet)}\ = 
p_{-1}(\beta)\, p_{-1}(\pt)^{d-1}\,{\bf 1} 
\ \in\ H_2(\hilb^d(K3),\IZ)\,,
\ee
is a curve class whose associated Gromov-Witten invariants we will consider later.
It is of partition type $\pi_\bet=[(1,\beta),(1,\pt)^{d-1})]$, where $\pt$ denotes the point class which is dual to $H^4(K3,\IQ)$; 
note that we implicitly sum here and in the following over cyclic permutations.
The curve $C(\beta)$ involves only one non-trivial length one piece, so it will eventually correspond to a string that starts and ends at the same copy of $K3$. 

The following class will be more interesting for us: it corresponds to the exceptional curve that arises as
blow-up of the orbifold singularity of $\sym^d(K3)$:
\be\label{Adef}
A \ \equiv \gamma_{(\pi_A)}\ =  
p_{-2}(\pt) p_{-1}(\pt)^{d-2}\,{\bf 1} 
\ \in\ H_2(\hilb^d(K3),\IZ)\,,\ \ d\geq2\,.
\ee
Its partition type is $\pi_A=[(2,\pt),(1,\pt)^{d-2})]$.
In physics language this would correspond to a string in the twisted sector that links two copies of $K3$.
We combine these two classes by writing
\be\label{Cdef}
C_{\bet,k}\equiv C(\bet)+ kA \in H_2(\hilb^d(K3),\IZ) \, ,\  \bet\in H_2(K3,\IZ)\,, k\in \IZ\,,
\ee
where we understand that $k\equiv0$ if $d=1$.

Moreover, we will need a few more cohomology classes. First, the
cohomology dual of $A$ is 
\be\label{sigma2}
\gamma^{(\pi_A)}=p_{-2}({\unit}) p_{-1}({\unit})^{d-2}{\unit}\in H^2(\hilb^d(K3),\IQ)\,, 
\ee 
which corresponds in CFT language to the universal $\IZ_2$ twist field,
$\sigma_2$,
that couples to the blow-up modulus of the orbifold singularity.
Here, $\unit\in H^{0}(K3,\IQ)$ denotes the unit.
Moreover we define
\be\label{gammaFd}
\gamma^{(\pi_{\alpha^d})}\ = \ p_{-1}(\alpha)^{d}\,{\unit}  \in H^{d,d}(\hilb^d(K3),\IQ)\,,
\ee
which is of partition type $\pi_{\alpha^d}=[(1,\alpha)^d]$ with K\"ahler class $\alpha\in H^{1,1}(K3,\IQ)$.
Finally we need the class
\be\label{gammaF}
\gamma^{(\pi_\alpha)}\ = \ p_{-1}(\alpha) p_{-1}(\unit)^{d-1}\,{\unit}  \in H^{1,1}(\hilb^d(K3),\IQ)\,,
\ee
with $\pi_\alpha=[(1,\alpha),(1,\unit)^{d-1}]$, which will correspond to a marginal K\"ahler deformation.

\subsection{Reduced Gromov-Witten invariants} \label{sec_reduced}

We now introduce Gromov-Witten invariants that in physics serve as building blocks for correlation functions.
Naively these are given by counting maps $\varphi: \Sigma_{g,n}\rightarrow X_N$ of genus $g$ curves with $n$ 
punctures into a given Calabi-Yau $N$-fold in question, which we keep generic for the moment.
One may in addition specify the homology class of the image, which we denote by $C = \varphi_*[\Sigma_{g,n}]\in H_2(X_N,\IZ)$. However what one really considers are integrals over the suitably compactified moduli spaces, $\overline\cM_{g,n}$, of stable  maps. More generally one considers {\it relative} Gromov-Witten invariants counting curves that pass through cycles dual to
given cohomology classes $\gamma_j\in H^{q_j q_j}(X_N,\IQ)$:
\be\label{GWintegral}
\langle \gamma_1\dots \gamma_n\rangle_{g,n;C}^{X_N}\ =\ \int _{[(\overline{\cM}_{g}(X_N,C),\{\gamma_j\}]^{vir}}ev_1^*(\gamma_1)\cup\dots ev_n^*(\gamma_n)\,.
\ee
Here $ev^*$ denotes the pullback by the evaluation map,
$ev\!:~\overline{\cM}_{g,n}(X_N,C) \rightarrow X_N$. 
In order for these invariants not to vanish identically, the expected, or virtual dimension, given by
\be\label{virtdim}
{\rm vdim}\,[\overline{\cM}_{g,n}(X_N,C),\{\gamma_j\}]^{vir}= (N-3)(1-g) -\sum (q_j-1)\,,
\ee
must vanish.

So far this is the well-known story designed for Calabi-Yau threefolds, for which ${N=3}$ so that the genus dependence 
of the virtual dimension disappears. However, in the present situation
where we deal with $2d$ complex-dimensional hyperk\"ahler manifolds, $X_{2d}=\hilb^d(K3)$, the story needs to be refined, because 
the GW invariants as defined above vanish identically. This can be understood in various ways. In physics this mirrors the fact that
due to the $N=(4,4)$ supersymmetry, non-renormalization theorems protect certain correlation functions.
Mathematically, the problem originates \cite{Bryan1997TheEG, Maul_0705.1653,Maulik:2010jw,Klemm_0807.2477,Pandharipande:2014ooa} from the fact that a generic, non-algebraic deformation moves a curve $C$ of Hodge type $(1,1)$ to one with a type $(2,0)$ or $(0,2)$ component, so that it is not holomorphic any more. Therefore it does not contribute to
the counting of holomorphic maps, and by deformation invariance the Gromov-Witten invariants must then vanish even for holomorphic curves of type $(1,1)$. 

Nevertheless, one can define generalized {\it reduced} GW invariants, 
abstractly speaking by modifying the obstruction theory such that the $(2,0)$ and $(0,2)$ components do not appear \cite{Maul_0705.1653,Maulik:2010jw}.
Concretely, one can construct reduced invariants by considering $X_{2d}=\hilb^d(K3)$  as fiber of some $(2d+1)$-fold \cite{Bryan1997TheEG, Maul_0705.1653,Maulik:2010jw,Klemm_0807.2477,Pandharipande:2014ooa}. This ensures that the fibral classes stay at Hodge type $(1,1)$ under algebraic deformations. 
There are various, basically equivalent possibilities to implement this, and in the present context
it is most natural to view $X_{2d}$ as fiber of its twistor family $T_{2d+1}\rightarrow \IP^1$, where $\IP^1$ is
stands for the unit sphere of complex structures on hyperk\"ahler manifolds.

The crucial point \cite{Bryan1997TheEG} 
is now that while for generic complex structure there are no holomorphic curves, there exists
a unique, fixed choice of complex structure for which holomorphic curves in class $[C]$ with $C^2\geq-2$ exist, so that they can be meaningfully counted in terms of non-vanishing invariants.
Moreover, the modified obstruction theory leads to a modified, reduced version of the virtual dimension,
\be\label{reduceddim}
{\rm vdim}\,[\overline{\cM}_{g,n}(\hilb^d(K3),C),\{\gamma_j\}]^{red}\ =\ (2d-3)(1-g)-\sum_{j=1}^n(q_j-1)+1\,,
\ee
which is one dimension larger than the unreduced one. This allows to abstractly define
non-vanishing reduced GW invariants for hyperk\"ahler Hilbert schemes with $d\geq1$
as follows:
\be\label{totalcorr}
\big\langle\,\gamma_1,\dots,\gamma_n\,\big\rangle_{g;C}^{\hilb^d(K3)}\ :=\
\int_{[(\overline{\cM}_{g,n}(\hilb^d(K3),C),\{\gamma_j\}]^{red}}ev_1^*(\gamma_1)\cup\dots ev_n^*(\gamma_k)\,.
\ee
In fact, most combinations of $d>1$ and $g>1$ still lead to vanishing invariants as these are genuinely zero (the exception being for $d=g=2$), but for physics only $g=0,1$ is relevant if $d>1$. In this paper we focus on $g=0$ only,
which corresponds to $\Sigma=\partial$AdS$_3=\IP^1$ on the gravity side.

\subsection{Interpretation in terms of the topological $N=4$ string} \label{sec_topstring}

We now suggest how the foregoing abstract mathematics can be translated 
to a more tangible physical language as follows. The key structure is
the topologically twisted $N=(4,4)$ superconformal symmetry of the sigma model on $\hilb^d(K3)$.
In the left-moving sector,
the relevant ``small'' $N=4$ superconformal algebra \cite{Eguchi:1987sm} 
is generated by the familiar holomorphic currents $J,T,G^\pm$ that form an $N=2$ subalgebra,
extended by the \mbox{(super-)}currents $J^{++},J^{--}$ and $\tilde G^\pm=\oint J^{\pm\pm}G^\mp$,
$G^\mp=-\oint J^{\mp\mp}\tilde G^\pm$.
Its central charge is given by $\hat c=2d$, which vanishes after topologically twisting.
The bosonic currents extend the $U(1)$ symmetry to $SU(2)$, under which the 
supercharges transform as two doublets. 
The same applies to the right-moving sector that we don't write
explicitly.

The $SU(2)$ symmetry reflects in CFT that in hyperk\"ahler geometry, there is no a priori notion of  holomorphicity. That is, any chiral primary field, satisfying
\be\label{chiralpri}
\hG^+_0 \phi \ =\ \htG^+_0 \phi\ =\  (L_0-\shalf J_0)\phi\ =\ 0\,,
\ee
with charge $q=1$, can be transformed to an equivalent anti-chiral field via
a local $SU(2)$ transformation,
\be\label{antichiral}
\int J^{--}\phi_j\ =:\ \sum_\bark{M_j}^\bark \bar\phi_\bark\,,
\ee
where $M$ is some unspecified numerical mixing matrix \cite{Berkovits:1994vy,Berkovits:1999im}.
Thus, when combining this with the right-moving counterparts, the (chiral,chiral) ring, which 
represents elements of $H^{1,1}(K3)$, becomes identified not only with
the (anti-chiral,anti-chiral) ring, but also with the twisted (anti-chiral,chiral) ring, which
formally represents elements of $H^{-1,1}(K3)$, as well with its complex conjugate. 
In geometry, this reflects the equivalence of holomorphic and complex structure
deformations, implemented by contraction with the holomorphic $(2,0)$-form.

A crucial new feature of $N=4$ superconformal theories
 is that there is an additional, ``global'' $\widehat{SU}(2)$, which acts as outer automorphism
by mixing the supercurrents of the same charge and so is compatible with the
``local'' $SU(2)$ symmetry generated by $J^{\pm\pm}$. It is however not a symmetry, and  
accordingly one can form the following family of
supercurrents~\cite{Berkovits:1994vy,Berkovits:1999im},
\bea\label{zetafamily}
\hG^+(\zeta) &=& \zeta_1\,G^+ + \zeta_2\,\tG^+\nn
\\
\hG^-(\zeta) &=& \zeta_1^*\,G^- + \zeta_2^*\,\tG^-
\\
\htG^-(\zeta) &=& -\zeta_2\,G^- + \zeta_1\,\tG^-\nn
\\
\htG^+(\zeta) &=& -\zeta_2^*\,G^+ + \zeta_1^*\,\tG^+\nn
\eea
for which the inequivalent choices are parametrized by $\widehat{SU}(2)/U(1)\simeq \IP^1_\zeta$ defined by $|\zeta_1|^2+|\zeta_2|^2=1$. It represents a family of inequivalent embeddings of the $N=2$ subalgebra 
spanned by $\widehat{G}^\pm(\zeta)$, within the full $N=4$ superconformal algebra.
Translated back into the mathematical terms of the foregoing section,
$\IP^1_\zeta$ corresponds to the unit sphere of possible complex structures within the hyperk\"ahler structure.
Fixing a point on $\IP^1_\zeta$ corresponds to picking a particular member
of the family of inequivalent $N=(2,2)$ subalgebras, relative to which one can define a notion of 
holomorphic, or primary chiral fields.\footnote{Much more could be said about these matters, but this would lead us too far away. We refer the reader to refs.~\cite{Berkovits:1994vy,Aspinwall:1996mn,Dijkgraaf:1998gf,Berkovits:1999im,deBoer:2008ss}
for more details.}

Concretely, in order to gain some intuition, consider  as an example a topologically twisted, $N=4$ superconformal sigma model on an algebraic $K3$ surface
(ie, one that is described by a holomorphic embedding via an algebraic equation $W=0$ 
in some weighted projected space). Writing down an equivalent
$N=2$ supersymmetric Landau-Ginzburg model with superpotential
$W$ gives a canonical definition of the $N=2$ supercharges, 
which can be identified with $\hG^\pm(\zeta)$  in ({\ref{zetafamily}) with $\zeta_1=1$, ${\zeta_2=0}$.
Algebraic curves embedded in this $K3$ then provide holomorphic representatives 
of Hodge type $(1,1)$ that are
aligned with this fixed complex structure. Moreover, algebraic deformations of $W$, 
which are characterized by the Picard lattice $Pic(K3):\equiv H^2(K3,\IZ)\cap H^{1,1}(K3)$, will preserve  
the holomorphicity of these curves~\cite{Aspinwall:1996mn,Dijkgraaf:1998gf}. Thus, 
restricting to algebraic surfaces and their deformations provides a way to
define reduced invariants and count holomorphic curves.

This brings us to correlation functions, 
for which we write only the left-moving part for notational simplicity.
The four supersymmetries impose constraints which tend to make correlation functions
vanish unless one modifies them in a suitable way. Our purpose here is to connect this modification
to the reduced invariants of the previous section.
The strategy is to adopt some ideas of $N=4$ topological strings \cite{Berkovits:1994vy,Berkovits:1999im}
and apply them to the $N=(4,4)$ superconformal sigma model on $\hilb^d(K3)$ under consideration.

For our application where $\Sigma=\IP^1$, only $g=0$ is relevant. 
The usual procedure to deal with vanishing genus is to consider $m$-point correlators with $m\geq 3$,
and cancel three of the $\hG_{-1}^-$ of the integrated insertions against the three negative powers
associated with the Beltrami differentials. This amounts to mark three points on the sphere
to fix its automorphisms. In the end, the modified, genus zero correlator of $N=4$ strings, as proposed
in \cite{Berkovits:1994vy,Berkovits:1999im}, takes the form
\be\label{F0m3}
\cF_{0,m\geq3}(\zeta)\ =\
\Big\langle-2d\,\,\Big\vert\,
\prod_{i=1}^3 \phi_j\,\big(\int\htG^+\big)^{-1}\!\int \!J\,\,\prod_{j=1}^{m-3}\int\hG_{-1}^-\phi_{j+3}
\Big\rangle^{\hilb^d(K3)}_{0,m}\,.
\ee
In the left bracket we have indicated the background $U(1)$ charge $q_0=\hat c( g-1)$,
which arises from the topological twist and which needs to be cancelled by the sum of the charges in the correlator. We also indicated the dependence on the $\IP^1_\zeta$-worth of complex structures
of the supercharges (\ref{zetafamily}).

The new ingredients, as compared to correlators of standard topological strings,
are the insertions of $(\htG^+)^{-1}$ and $J$. They are crucial for making
non-vanishing correlators of the desired kind possible.
In particular, the first insertion implements the modified charge selection rule  (\ref{reduceddim})
of the reduced Gromov-Witten invariants (\ref{totalcorr}). That is,
if we take for the remaining integrated insertions moduli fields with $q_{j+3}=1$, the condition on the
unintegrated insertions is $\sum _{i=1}^3 q_j= 2d+1$, just like for a Calabi-Yau $(2d+1)$-fold. 
However, one needs to make sense of the formal inverse power of $\htG_0^+$
(which, when adopting a field theoretical language, one may view as propagator of the form
$\htG_0^-/ \{\htG_0^+,\htG_0^-\}$).
There are at least two, essentially equivalent ways to deal with that.

First, in many instances the cohomology of $\htG_0^+$ is trivial; this applies to the examples for topological $N=4$ strings
of refs.~\cite{Berkovits:1994vy,Berkovits:1999im}, which include NSR strings on $K3$.  
This means that we can pick one of the un-integrated insertions and write it as $\phi_j=\int\htG^+ V_j$, where
$V_j$ has a $U(1)$ charge reduced by one. Then $\int\htG^+$ can be moved to $J$, which gives
it back and in the second step one can cancel its inverse power.  In the end the genus zero
correlator takes the form
\be\label{threepoint}
\cF_{0,m\geq 3}(\zeta)\ =\
\Big\langle\!-2d\,\,\Big\vert\,
\phi_1 \phi_2 V_3\, \prod_{j=1}^{m-3}\int\hG_{-1}^-\phi_{j+3}\Big\rangle^{\hilb^d(K3)}_{0,m}\,.
\ee
If we were to consider topological $N=4$ strings in NSR formulation, we can take
$\zeta_1=1$, ${\zeta_2=0}$ in ({\ref{zetafamily}), and $\htG^+=\eta$, so that
it can be easily inverted:
$(\htG^+)^{-1}=\xi$. Here $\eta$ and $\xi$ are fermions with charges $q=\pm1$ that arise from
bosonizing the  familiar $(\beta,\gamma)$ super-ghost system. Hence $V_3=\xi\phi_3$ and the
correlator takes the form $\langle\phi\phi\phi\xi_0\prod\int G^- \phi\rangle$.
Thus the zero mode of $\xi$ not only ensures the correct charge balance, but also provides
a modified,  enlarged Hilbert space along the lines of ref.~\cite{FRIEDAN198693}.
We expect this interpretation of the formal insertion $(\htG^+)^{-1}$ to be morally valid also beyond this
suggestive example. 

Alternatively, if $m\geq4$ one can make use of the local $SU(2)$ symmetry transformations
(\ref{antichiral}) and $\hG^-=\int J^{--}\htG^+$, which together imply:
\be\label{swap}
\hG_{-1}^- \phi_j \ =\ \htG_0^+ \,\sum_\bark{M_j}^\bark \bar\phi_\bark \,.
\ee
Picking an integrated insertion and transforming it into an anti-chiral field,
we can move as before the contour of $\int \htG^+$ first on $J$ and then on top of its inverse to cancel it. This leads to
\be\label{noholcorr}
\cF_{0,m\geq4}(\zeta)\ =\
\Big\langle\!-2d\,\,\Big\vert\,
\phi_1 \phi_2 \phi_3\sum_\bark  {M_4}^\bark\! \int \bar\phi_\bark  \, \prod_{j=1}^{m-4}\int\hG_{-1}^-\phi_{j+4}\Big\rangle^{\hilb^d(K3)}_{0,m}\,.
\ee
This can also be directly obtained from (\ref{threepoint}) by applying (\ref{swap}) if $m\geq4$.
Note that in the twisted theory, $\bar\phi$ has conformal dimension equal to one
and so can be integrated over. Being charged it is not an ordinary, symmetry preserving
 modulus, but rather its role is to provide the correct charge balance for the reduced invariant.

We thus see that the inverse power of $\htG_0^+$ is not only crucial for obtaining non-vanishing
correlators in the first place, but also for obtaining non-vanishing deformations by marginal operators.
The non-renormalization theorems of refs.~\cite{deBoer:2008ss,Baggio:2012rr}  apply when this insertion is not there:
the argument uses the symmetry transformation (\ref{swap}) to map a chiral field to an equivalent
anti-chiral one, and moving $\int \htG_0^+$ to the three un-integrated insertions yields zero by the
physical state condition (\ref{antichiral}). 
More specifically, refs.~\cite{deBoer:2008ss,Baggio:2012rr} consider the
untwisted theory which means that the background charge is carried by an anti-chiral field
with charge $-2d$. Thus altogether two anti-chiral fields appear in the correlator and for this configuration
the non-renormalization theorem does not apply.

Note that both correlators (\ref{threepoint}) and (\ref{noholcorr}) can be defined
just within conformal field theory, which is what is relevant at the boundary of AdS$_3$. 
This lets us to propose, tentatively,  the following correspondence between the reduced invariants (\ref{totalcorr})
and topologically twisted correlation functions:
\be\label{redcorrdef}
\big\langle\,\gamma_1,\dots,\gamma_n\,\big\rangle_{0,n}^{\hilb^d(K3)}
\ \simeq\
\Big\langle-2d\,\,\Big\vert\,
\prod_{i=1}^n \gamma_j\,\big(\int\htG^+\big)^{-1}\,
\Big\rangle^{\hilb^d(K3)}_{0,n}\,,
\ee
where we did not write possible marginal deformations explicitly and where $\htG^+$ refers to the
corresponding supercharge of the CFT sigma model on $\hilb^d(K3)$.
Note that here the $n$ relative insertions $\gamma_j$  are not integrated over.
Thus, the relevant selection rule for general $n$ is given by 
\be\label{lhsconstr}
 2d+1\ \stackrel{!}{=}\
\sum_{j=1}^n q^{{(j)}}\,,
\ee
where $q^{{(j)}}$ are the charges (\ref{degree}) of the $\gamma_j$.
Moreover also note, as discussed in Section~\ref{sec_reduced}, that the definition of reduced invariants
implicitly involves an appropriate choice of complex structure in order to be nonzero, and this
corresponds to a definite choice of the $\zeta_i$ for the family of supercharges (\ref{zetafamily}).

Concluding, we have outlined how the mathematical notion of reduced Gromov-Witten invariants
could be translated into the physical language of topological $N=(4,4)$ strings,
which employ the specific definition (\ref{F0m3}) of correlation functions  \cite{Berkovits:1994vy,Berkovits:1999im}.
This definition was designed to generate non-vanishing correlation functions,
in a similar spirit as reduced GW invariants of hyperk\"ahler manifolds were introduced as a modification of ordinary ones.
While it would be certainly desirable to have a better understanding of this tentative connection,
the following considerations are largely independent from it; the aim of this section
was to motivate the reduced invariants from a physical perspective.

The main question is clearly whether the prescription (\ref{redcorrdef}) on the CFT 
side leads to physically meaningful correlation functions. Encouraging is that
the exact free-field formulation of strings on the AdS side, as developed in \cite{Dei:2020zui,Bhat:2021dez,Gaberdiel:2022als,Knighton:2022ipy,McStay:2023thk}, 
does realize a topological $N=4$ string, which suggests that
considering reduced invariants is natural. 
See however the concluding Section~\ref{sec_summary} for some important caveats.

For now we adopt the relevance of reduced invariants 
as a working hypothesis and leave a more detailed analysis for later work. 
With this in mind we now proceed to the central theme of this paper.

\vfil

\section{The Gromov-Witten/Hilbert Correspondence} \label{sec_GWH}

The link between the ``Gromov-Witten" and ``Hilbert'' boxes in Figure \ref{fig_correspondences} 
can now be given a concrete meaning in terms of the following key formula.\footnote{
While we focus here on the Hilb/GW Correspondence for 
$\Sigma=\partial$AdS$_3=\IP^1$, let us note that for thermal AdS$_3$ with torus boundary, $\Sigma=\IE$,
there is a noteworthy difference in that the relation between invariants of the 
Hilbert and Gromov-Witten sides has an extra term \cite{Oberdieck:2014aga,Oberdieck:2021ftp}
as compared to eq.~(\ref{GWhilb}). In a broader context, it has the interpretation of a wall crossing term
\cite{Nest_2111.11417,Nest_2111.11425,Nest_2208.00889}. Such wall crossing is absent for $\Sigma=\IP^1$ that we consider here.}
It has been derived and discussed in refs.~\cite{Ober_1406.1139,Oberdieck:2014aga,Ober_1605.05238,Oberdieck:2022khj,Genlik_2304.06536v1}
where it has been proven for $n\leq3$ and conjectured to be more generally valid.
For $d\geq1$ and assuming that  $\bet$ be a primitive\footnote{Multicover contributions
are fully determined by Hecke transforms \cite{Oberdieck:2021kgf,Oberdieck:2022khj} and thus do not provide independent extra
information.}
and effective class, we write it in the following form:
\be\label{GWhilb}
 \sum_{k\in\IZ} \,y^k\langle\,\gamma_1,\dots,\gamma_n\,\rangle_{0;C_{\bet,k}}^{\hilb^d(K3)} 
\ =\ 
u^{N_0(d,\mu)}\sum_{g\geq0}u^{2g-2}\
\langle\,\gamma_1,\dots,\gamma_n\,\rangle_{g;(\bet,d)}^{K3\times \IP^1/\{z_1,\dots,z_n\}}\,, 
\ee
where, as will be explained later: $N_0(d,\mu):=2d-\sum_j^n\sum_i^{\ell(\mu^{(j)})}(\mu_i^{(j)} -1)$, and 
the brackets stand for connected, reduced invariants (\ref{totalcorr}).
Here we use the same symbol, $\gamma_i$, for insertions of primary chiral fields on both sides,
because they encode equivalent mathematical data despite representing cohomology
classes on different spaces.

Moreover, the map between the expansion parameters on both sides is given by
\be\label{shift}
y\ =\ e^{2\pi  i(z+1/2)}\ \equiv\ - e^{-u}\,.
\ee
Thus the identity 
eq.~(\ref{GWhilb}) amounts to a non-trivial resummation between two a priori very different expansions, namely an 
exponential, instanton type expansion in the blow-up modulus of $\sym^d(K3)$ on the left hand side,
 and a power type, perturbative  genus expansion on the right. 
 This is analogous to the Donaldson-Thomas/Gromov-Witten correspondences as discussed for example in refs.~\cite{Maulik2003GromovWittenTA,Maulik2004GromovWittenTA, Pandharipande:2011jz}.
In fact (\ref{GWhilb}) is a special example within a much more general framework of correspondences as discussed 
in~\cite{Nest_2111.11417,Nest_2111.11425,Nest_2208.00889}.

A generic feature is the symmetry $y\leftrightarrow 1/y$ which is not manifest in a series or Laurent expansion in $y$ or $/1/y$.
Also note that $u$ is taken here to be complex, and its imaginary part
plays the role of a theta-angle. The shift of origin in  (\ref{shift}) reflects a 
constant $B$-field background in the sigma-model on $\hilb^d(K3)$
at the orbifold point in the moduli space, which is a familiar phenomenon in string geometry. 
On the AdS-side the role of the theta-angle is played by a linear combination of RR zero-and four-forms, $C_0$ and $C_4$,
see for example \cite{Larsen:1999uk,OhlssonSax:2018hgc}.

Let us now focus on the left- and right-hand sides of the formula  eq.~(\ref{GWhilb}) separately, 
for more mathematical details see eg.~\cite{Ober_1406.1139,Oberdieck:2014aga,Ober_1605.05238,Oberdieck:2022khj,Genlik_2304.06536v1}.

\subsection{The Hilbert side} \label{sec_LHS}

The ingredients of the left-hand side of (\ref{GWhilb}) 
have been extensively discussed in the preceding sections: the reduced GW invariants count holomorphic
maps of genus zero curves into the
class $C_{\bet,k}\equiv \ C(\bet)+k A\in H_2(\hilb^d(K3),\IZ)$, subject to incidence conditions imposed by the cohomology
elements \mbox{$\gamma_j\equiv\gamma^{(\pi^{(j)})}\in H^{*,*}(\hilb^d(K3),\IQ)$} (which just means that the curves
intersect cycles Poincar\'e dual to the $\gamma_j$).  Recall that these cohomology elements are defined in terms of 
 cohomology weighted partitions $\pi^{{(j)}}=[(\mu_i^{{(j)}},\al_i^{{(j)}})]$  of size $d=\sum\mu_i$.  
Recall also that the condition (\ref{lhsconstr}) for non-vanishing, reduced invariants (\ref{redcorrdef}) reads
\be\label{lhsconstrdef}
 2d+1\ \stackrel{!}{=}\
\sum_{j=1}^nq^{{(j)}}+ \sum_{j=n+1}^m(q^{{(j)}}-1)\,.
\ee
if we add extra integrated insertions. Clearly marginal operators
based on primary chiral fields with $q^{{(j)}}=1$, $j=n+1,...m$, do not contribute.

Let us focus on the dependence on the most interesting of such marginal deformations, 
namely the blow-up mode $u$, which we keep here real for simplicity.
It parametrizes the deformation by the marginal twist field,
\be\label{marginal}
\Big\langle\gamma_1,...,\gamma_n\,\exp{\big(\!-u \int_{\IP^1}\gamma^{(\pi_A)[2]}\big)}\,\Big\rangle_{0;C_{\bet,k}}^{\hilb^d(K3)}\,,
\ee
where $\gamma^{(\pi_A)[2]}\equiv \hG_{-1}^-\bar \hG{}_{-1}^-\gamma^{(\pi_A)}$ and 
$\gamma^{(\pi_A)}\in  H^{1,1}(\hilb^d(K3),\IQ)$ is a 
holomorphic representative of  the two-form defined in (\ref{sigma2}). 
It represents the K\"ahler form of the exceptional curve $A$ (\ref{Adef}), 
such that $u=0$ amounts to zero volume. 
It appears on the left-hand side of (\ref{GWhilb}) exponentiated via $y=-e^{-u}$, 
and the powers of~$y$ count holomorphic maps from $\IP^1$ into the curve $kA$. 
In the language of the topological sigma model on $\hilb^d(K3)$, such maps correspond to
holomorphic world-sheet instantons whose action is proportional to the volume $u$ of the exceptional curve.
This connection can be made more explicit by first expanding (\ref{marginal}) and then writing:
\bea
\sum_m \frac {(-u)^m}{m!}\Big\langle\prod_{j=1}^n \gamma_j\big(\int_{\IP^1} \gamma^{(\pi_A)[2]}\big)^m\Big\rangle_{0;C_{\bet,k}}^{\hilb^d(K3)}
\!\!\!\!\!& = &\!\!\!
\sum_m \frac {(-u)^m}{m!}\big(\int_{C_{\beta,k}}\gamma^{(\pi_A)} \big)^m \big\langle\prod_{j=1}^n \gamma_j\big\rangle_{0;C_{\bet,k}}^{\hilb^d(K3)} 
\nn
\\
 &=&\!\!\!    e^{-u\int_{k A}\gamma^{(\pi_A)}}\big\langle\gamma_1,...,\gamma_n\big\rangle_{0;C_{\bet,k}}^{\hilb^d(K3)}  
 \\
& =&\!
 (-1)^k {y^k} \,\big\langle\gamma_1,...,\gamma_n\big\rangle_{0;C_{\bet,k}}^{\hilb^d(K3)}  \,.          
\nn                                                                      
\eea
In the first line we used the divisor axiom \cite{Hori:2003ic}, 
the second line uses $C_{\bet,k}\equiv C(\bet)+ kA$ as per (\ref{Cdef}), 
and the third $\int_A\gamma^{(\pi_A)}=1$. 
Analogous arguments hold for holomorphic K\"ahler moduli of $K3$ as defined
in (\ref{gammaF}).

Note that in theories with $N=(2,2)$ supersymmetry, such expansions are a priori
well defined for large K\"ahler moduli only, 
which corresponds to a weakly coupled geometric sigma model regime.
In the present situation with $N=(4,4)$ supersymmetry there is no obstacle against tuning the blow-up modulus
to $u=0$, and in fact it will turn out that the expansion is indeed best defined near $u=0$. 
This brings us to the right-hand side of the correspondence~(\ref{GWhilb}).

\subsection{The Gromov-Witten side} \label{sec_RHS}

The right-hand  side of (\ref{GWhilb}) is structurally more complicated. 
Here we sum over relative, reduced Gromov-Witten invariants
counting genus $g$ maps into class $(\bet,d)$, where $\bet\in H_2(K3,\IZ)$ as before
and $d$ is the degree of the map to $\IP^1$. 
Imposed on these maps are incidence conditions implemented
 by the cohomology elements  \mbox{$\gamma_j\equiv\gamma^{(\pi^{(j)})}
 \in H^{*,*}(K3\times \IP^1/\{z_j\})$}, which are tied to fixed positions  $\{z_j\}$
on~$\IP^1$. 
The partition types $\mu^{(j)}$ on the Hilbert side translate
into ramification profiles over $\{z_j\}$ of the same types, now pertaining to the $d$-fold cover of $\IP^1$.
This is in line with the familiar viewpoint from AdS/CFT, 
where one counts covering maps from the world-sheet to $\IP^1=\partial{\rm AdS}_3$ 
with the incidence conditions that they pass through the preimages of the branch points  $\{z_j\}$.
Indeed this is precisely what is meant by relative invariants, as far as $\IP^1$ is concerned.
  
More specifically, for a given cohomology weighted partition $\pi^{(j)}=[(\mu_i^{(j)},\alpha_i^{(j)})]$, $i=1,..,\ell(\pi^{(j)})$, we have an insertion of $\prod_i\alpha_i^{(j)}$, where each $\alpha_i^{(j)}\in H^{*,*}(K3,\IQ)$ is tied to the corresponding cycle over $z_j$ that links $\mu_i^{(j)}$ sheets together. 
One should consider this ``product" as short-hand for repeated entries.
Thus, if individual partitions involve multiple cycles, this may correspond to higher than $n$-point functions (which
are possibly, but not necessarily disconnected -- see Figure~\ref{fig:weightedpart} for an example). 
Putting this together, we can thus write explicit representations of the 
$\gamma_j$ on the right-hand side of (\ref{GWhilb})
in term of twist fields $\sigma_{\mu_i^{(j)}}$ as follows:
\be\label{sigmapi}
\gamma_j\ \equiv\ \gamma^{(\mu^{(j)},\alpha^{(j)})}
\ =\  \prod_{i=1}^{\ell(\mu^{(j)})} (\sigma_{\mu_i^{(j)}}\alpha_i^{(j)})
\ee
(for notational convenience, we dropped here the symmetry factor $\mathfrak{z}(\pi)$).
The twist fields $\sigma_{\mu_i^{(j)}}$ are the analogs of the
Nakajima operators, $p_{-\mu_i^{(j)}}$ in (\ref{nakaj}), and create branch points of
order $\mu_i^{(j)}$ at the locations $z_j$ on $\IP^1$. In (\ref{sigmapi}) we indicated
by the parentheses that the insertions of the twist fields and of the $\alpha_i$ are
correlated and tied to particular sheets if there are multiple cycles over a given branch point.

\begin{figure}[t!]
\centering
\includegraphics[width=9cm]{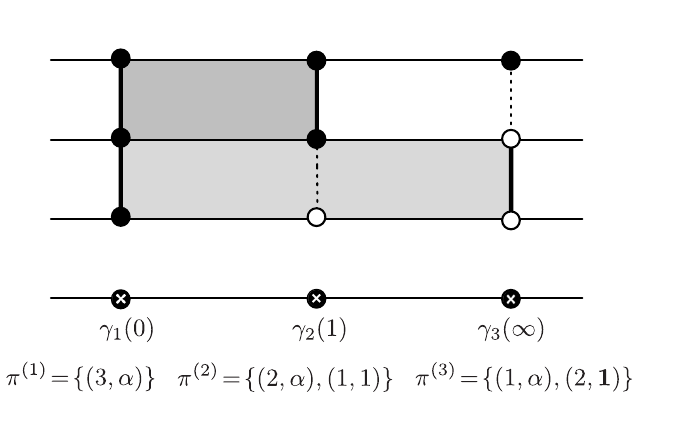}
\caption{Cohomology weighted partitions $\pi^{(j)}=\{(\mu_i^{(j)},\al_i^{(j)})\}$ defining
a $d=3$ sheeted covering of $\IP^1$ (horizontal lines).
Solid vertical lines link $\mu_i$ sheets at the ramification points, dashed lines indicate unramified coverings,
and the shaded areas signal branch cuts that may be viewed as bridges between the sheets. 
Solid dots signify insertions of  $\alpha\in H^{1,1}(K3)$, while open dots indicate trivial insertions. The three shown ramification profiles with in total 4 non-trivial cycles correspond to the connected genus zero invariant 
$\langle\,\sigma_3(\al)\sigma_2(\al)\sigma_1(\al)\sigma_2(1)\,
\rangle_{0}^{K3\times \IP^1/\{0,1,\infty\}}$, 
where the twists $\sigma_{\mu_i}$ are as defined in (\ref{sigmapi}) and we denote trivial twists by $\sigma_1$.
In physical language, the first two insertions correspond to massive, ``spectrally flowed'' K\"ahler classes, while the
last two correspond to one massless K\"ahler modulus plus one massless orbifold blow-up modulus.
}
\label{fig:weightedpart}
\end{figure}

A priori, the charge selection rule for non-zero correlators, resp.~the reduced virtual dimension
of the moduli space, looks different for the non-Calabi-Yau threefold $K3\times \IP^1$,
for which the genus dependence drops. However,  there is an extra contribution of $2d$ from 
the non-zero first Chern class so that:
\bea\label{P1K3selrule}
{\rm vdim}[\overline{\cM}_{g,n}(K3\times \IP^1,\!(\bet,d)),\{\gamma_j\}]^{red}\!\! &=&\!
\left({\rm dim}_\IC(K3\times\IP^1)-3\right)(1-g)+\int_\bet c_1 (\IP^1)+1-\sum_{j=1}^n q^{(j)}
\nn\\
& =& \!2d+1-\sum_{j=1}^n q^{(j)}
\ \stackrel{!}{=}\ 0\,.
\eea
This reproduces the charge selection rule (\ref{lhsconstr}) for reduced invariants on the ``Hilbert''~side.

Let us momentarily focus on maps to branched coverings of $\IP^1$ for some
given degree $d$ and ramification profile $\{\mu^{(j)}\}$,  the
$K3$ going along for the ride. This data defines via the Riemann-Hurwitz formula
a curve $\Sigma^{(0)}$ with arithmetic genus
\be\label{gzerodef}
g_0\ =\ 1-d +\frac 12\Delta(\mu)\,,
\ee
where
\be\label{Deltadef}
\Delta(\mu) \ :=\ \sum_{j=1}^n\sum_{i=1}^{\ell(\pi^{(j)})} (\mu_i^{(j)}-1) \,.
\ee
We call $\Sigma^{(0)}$ the ``bare'' curve as it is defined for the unperturbed theory at $u=0$.
Depending on  $d$ and $\{\mu^{(j)}\}$, its arithmetic genus can be negative, in which case the
curve is necessarily disconnected. For example, in the completely unramified case where
all $\mu_i^{(j)}=1$ so that $\Delta(\mu) =0$, $\Sigma^{(0)}$  consists of $d$ disconnected curves $C_i\sim\IP^1$ with total 
 genus $g_0=\sum (g(C_i)-1)+1=1-d$. We will come back to this example in more detail in Section~\ref{sec_example}.

Since we aim for connected invariants for arbitrary $g\geq0$,
a certain number \mbox{$N(g,d,\mu)\geq0$} of extra insertions of the twist field
$\sigma_2$ will be generally needed in order to obtain non-vanishing invariants.
Morally speaking, in the language of conformal field theory we have
\be\label{K3P1corrg}
\Big\langle\,\big(\int_{\IP^1}\sigma_2^{[2]}\big)^{N(g,d,\mu)}\,
\prod_{j=1}^n \big( \prod_{i=1}^{\ell(\mu^{(j)})} (\sigma_{\mu_i^{(j)}}\alpha_i^{(j)})\big)(z_j)
\,
\Big\rangle^{K3\times \IP^1/\{z_j\}}_{g;(\bet,d)}\,,
\ee
where $\sigma_2^{[2]}\equiv \hG_{-1}^-\bar \hG{}_{-1}^-\sigma_2$  and where
we suppressed any other insertions; we will present an abstract, but precise mathematical expression momentarily. Importantly,
\bea\label{Ndef}
N(g,d,\mu)  \ :=\ 2(g-g_0)&=&  2g-2+2d -\Delta(\mu) 
\\
&=& 2g-2+N_0(d,\mu)\,,\nn
\eea
is the genus deficit between the bare curve $\Sigma^{(0)}$ and the genus $g$ curves we aim for.
Remembering that each power of the twist insertion
comes with a factor of $u$, this reproduces
the $u$-dependence on the right-hand side of (\ref{GWhilb}).

So far we have neglected $K3$. In general, if we count maps from
the world-sheet to $K3\times\IP^1$, there will be extra contributions from maps into $K3$,
which can be non-zero because we consider reduced GW invariants.  Thus, the genus of the world-sheet will
distribute over both $K3$ and ramified degree $d$ covers of $\IP^1$. Non-trivial maps to curves $\beta$ on $K3$ will be weighted by exponentiated K\"ahler moduli. There can also be degenerate, degree zero maps to points
on $K3$, an example of which we will meet at the end of Section~\ref{sec_perturbation}.

In (\ref{K3P1corrg}) we were writing down just a schematic expression for the Gromov-Witten invariant,
but there is a mathematically precise way of writing it \cite{Ober_1605.05238}, namely  as an integral over the
relative, reduced moduli space:
\be\label{GWK3P1}
\langle\,\gamma_1,\dots,\gamma_n\,\rangle_{g;(\bet,d)}^{K3\times \IP^1/\{z_j\}}
\ =\
\int_{[\overline{\cM}_{g,n}(K3\times \IP^1,(\bet,d)),\{\gamma_j\}]^{red}}
\sum_{l_j\geq0\atop \sum \l_j= N(d,g,\mu)}\prod_{j=1}^n \left(\psi_j^{l_j}\cup \prod_{i=1}^{\ell(\mu^{(j)})}ev_i^{*(j)} (\alpha_i^{(j)})\right),
\ee
up to symmetry factors.  The new ingredient is the Psi-class,
which is the first Chern class of the cotangent line bundle at the $i$-th point,
$\psi_i=c_1(\IL_i)$. More specifically,
the canonical interpretation of these invariants is in terms of topological
gravity \cite{Witten:1990hr,Verlinde:1990ku}, which provides a connection between Gromov-Witten and Hurwitz theory \cite{OkouPandha0204305}
(we will make use of this connection later in Section~\ref{sec_perturbation}).
In order not to disrupt the current flow of arguments, we delegated a few remarks to Appendix~\ref{app_topgr}.
We just mention here that in this formulation, the role of the twist fields is played by gravitational
descendants:\footnote{A connection between twist fields in the AdS/CFT Correspondence and gravitational descendants has been emphasized early in ref.~\cite{Rastelli:2005xaa}. } $\sigma_{\mu_i}\simeq \tau_{\mu_i-1}$, and in particular,
the insertions of $\sigma_2$ in (\ref{K3P1corrg}) correspond to in total $N(d,g,\mu)$ insertions of $\tau_1$.
The Psi-class in (\ref{GWK3P1}) is nothing but the representative of $\tau_1$ under the integral, distributed in
all ways over the $n$ branch points. 

Evidently, the question arises how to actually evaluate abstract integrals such as above
for concrete examples. Generically they can be expressed in terms of
complicated, ELSV-type Hodge integrals \cite{Ekedahl_1999,Ekedahl_2001}, which are hard to solve. However, using localization
and degeneration methods, explicit results have been obtained  for a variety of examples  in \mbox{refs.~\cite{Ober_1406.1139,Oberdieck:2014aga,Ober_1605.05238,Oberdieck:2022khj}},
and in Section~\ref{sec_example} we  make good use of one of these when making contact with the AdS perspective.

 \subsection{'t Hooft Expansion} \label{sec_tHooft}

Note that depending on the ramification profiles $\{\mu^{(j)}\}$ of the insertions $\gamma_j$, non-zero connected 
invariants for genus $g\geq0$ curves can arise also without extra insertions of the twist field,
ie., $N(g,d,\mu)=0$ for particular combinations of $g,d$ and $\{\mu^{(j)}\}$.
This is the ``undeformed'' situation at $u=0$ most often discussed in the physics literature, for which only a finite number of 
possible bare curves $\Sigma^{(0)}$ can contribute for given $d$, ie., $g_0$ is bounded from above \cite{Lunin:2000yv,Pakman:2009zz,Eberhardt:2019ywk,Eberhardt:2020akk}.
Deforming the theory by switching on $u$  can thus have two effects:  namely connecting initially disconnected sheets to produce
non-vanishing connected invariants, and/or inflicting higher genus corrections to $\Sigma^{(0)}$ to arbitrarily high orders.

Besides $u$ we have in general also K\"ahler moduli from the $K3$.
To be more specific, let us specialize to curves $\beta=\beta_h\in\ H_2(K3,\IZ)$ with self-intersection $\langle\beta_h,\beta_h\rangle=2h-2$. This 
is canonical for elliptic  $K3$'s where we can take $\beta_h=b+h\,f$, with $b,f$ being base and fiber curve classes, respectively 
(generic $K3$'s can be always deformed to this situation).
Denoting $q=e^{2\pi i \tau}$,  where $\tau$ is the complexified K\"ahler parameter  of the elliptic fiber, 
we can then sum also over $h$ and turn the RHS of (\ref{GWhilb}) into the following generating function:
\bea\label{Ftau}
\cF^{(d)}_{\{\pi^{(j)}\}}(q,u)&=& \sum_{g,h\geq0} u^{N(g,d,\mu) } q^{h-1}\, \GW_{g,\beta_h,\pi}
\\
&&\qquad {\rm where\ \ } \GW_{g,\beta_h,\pi}\ :\ =\  
 \langle\,\gamma_1,\dots,\gamma_n\,\rangle_{g;(\bet,d)}^{K3\times \IP^1/\{z_j\}}\,.\nn
\eea
This may be viewed as a generating function in the superselection sector 
specified by a fixed background geometry $\{\pi^{(j)}\}$, or bare curve $\Sigma^{(0)}$ 
(possibly with extra operator insertions such as shown in Figure~\ref{fig:weightedpart}), around which we expand  in $q$ and~$u$.  

Let us make the proposed connection between the GW/Hilb and the AdS/CFT Correspondences more concrete, by  addressing how the blow-up modulus $u$ relates to the string coupling constant on the AdS side.
They can't be identical, because, as said above, one can have background geometries $\{\pi^{(j)}\}$
with non-trivial correlators even at the orbifold point, $u=0$. These couple to
the six dimensional string coupling $g_s$ in the usual way.
Let us therefore weigh the generating function (\ref{Ftau}) by the appropriate power of $g_s$ and write
\bea\label{Nsubst}
{g_s}^{2g_0-2+ p} \cF^{(d)}_{\{\pi^{(j)}\}}(q,u) \nn
&=&  
\sum_{g\geq g_0^*,h\geq0} {g_s}^{2g_0-2+p} \, u^{N(g,d,\mu) } \, q^{h-1} \, \GW_{g,\beta_h,\pi}\nn\\
&=&
\sum_{g\geq g_0^*,h\geq0} {g_s}^{2g_0-2+p} \,   u^{2g-2g_0}\, q^{h-1}\,  \GW_{g,\beta_h,\pi}\\
&=&
\sum_{g\geq g_0^*,h\geq0} N^{1-g-\frac12 p}\, \lambda^{2g-2g_0}\,  q^{h-1}\,  \GW_{g,\beta_h,\pi}\nn\\
&=:& 
\cF^{(N)}_{\{\pi^{(j)}\}}(q,\lambda)\,.\nn
\eea
with $g_0^*=\max\{0,g_0\}$. If $N(g,d,\mu) = 2g-2g_0<0$, the degree constraint cannot be satisfied and
the Gromov-Witten invariant must vanish.  Moreover, if the bare curve is
disconnected so that $g_0<0$, $g$ starts at zero since we consider connected invariants only.
This is why we wrote the genus expansion as starting at $g=g_0^*$. 
Finally, $p=n d-\Delta$ denotes the number of cycles, or external legs.

In the third line we made the following substitutions: 
\be\label{AdSmap}
d \ =\ N\,,\ \ \
g_s \ =\ \frac 1{\sqrt N}\,, \ \ \ 
u\ =\ \frac{ \lambda}{\sqrt N}\,,
\ee
where $\lambda$ plays the r\^ole of a 't Hooft parameter which is taken to be constant as $N\rightarrow \infty$. 
Thus $\cF^{(N)}_{\{\pi^{(j)}\}}(q,\lambda)$ has the correct asymptotics for large $N$ \cite{Pakman:2009zz}. The new feature is the extra dependence on  $\lambda$ in terms of GW invariants relative to the bare curve $\Sigma^{(0)}$: switching on
$\lambda$ adds arbitrarily many extra handles to $\Sigma^{(0)}$.

 Note that there is such a 't Hooft expansion on top of each possible bare curve;
 even initially disconnected ones, as will be exemplified in the next section. 
 Each corresponds to a choice of fixed background geometry determined
by a collection ${\{\pi^{(j)}\}}$ of partitions of $N$. 
One may sum over such partitions, or bare curves for given $N$,
and for $N\rightarrow \infty$ this turns into a double genus expansion 
in both the string and 't~Hooft couplings, the latter being subleading in $1/N$.

 \section{A basic example} \label{sec_example}

So far we have discussed the formal structure of generating functions with generic ramification profiles.
We now focus on the simplest possible example, which is also the most significant one,
since the expansion starts from the most degenerate configuration.
The practical bonus is that this setup been already discussed in the mathematical literature \cite{Ober_1406.1139,Oberdieck:2014aga,Ober_1605.05238}. 

For this we consider again an elliptic $K3$ whose fiber and base classes we denote by $f,b\in H_2(K3,\IZ)$,
respectively, and count curves in the primitive class
\be
\bet_h\ =\ b+ h\,f\,,\qquad\quad \langle \bet_h,\bet_h\rangle\ =\ 2h-2\,,\quad h\geq0\,.
\ee
Concretely we will consider invariants (\ref{GWhilb}) involving two insertions of
$\gamma^{(\pi_{\alpha^d})}$ and one of $\gamma^{(\pi_{\alpha})}$ as defined in eqs.~(\ref{gammaFd}) and (\ref{gammaF}),
for $\alpha=f$.
These cohomology weighted partitions correspond to trivial ramifications. 
Thus all branched coverings will arise from 
insertions of twist fields $\sigma_2$, so that the genus expansion is not irritated
 by initial ramifications over $\{z_j\}=\{0,1,\infty\}$. See Figure~\ref{fig:unramified2}
and the next sub-section.

\begin{figure}[t!]
\centering
\includegraphics[width=7cm]{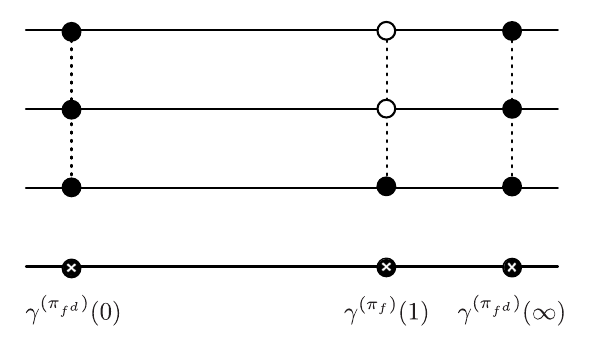}
\caption{Maximally degenerate, unramified bare geometry $\{\pi_{f^d},\pi_{f},\pi_{f^d}\}$ for $d=3$,
with notation as for Fig.~\ref{fig:weightedpart}. 
 It corresponds to a disconnected bare curve $\Sigma^{(0)}$ of arithmetic genus $g=1-d=-2$.
It represents three separate string world-sheets at the tensionless point $u=0$ in moduli space, with in total $2d+1=7$ insertions of massless K\"ahler moduli.
}
\label{fig:unramified2}
\end{figure}

The important point is that the connected generating function (\ref{Ftau})
\be\label{Fqu}
\cF^{(d)}_{\{\pi_{f^d},\pi_{f^{{\phantom |\!\!}}},\pi_{f^d}\}}(q,u) 
\ =\ 
u^{2d-2} \sum_{g,h\geq0} \,u^{2g}\, q^{h-1}
\Big\langle\,
\gamma^{(\pi_{f^d})}\,\gamma^{(\pi_{f})}\,\gamma^{(\pi_{f^d})}\,
 \Big\rangle_{g;(\beta_h,d)}^{K3\times \IP^1/\{0,1,\infty\}}\,,
\ee
can be determined exactly to all orders by localization and degeneration arguments \cite{Ober_1406.1139,Oberdieck:2014aga,Ober_1605.05238}, and is given in closed form~by:
\be\label{FOb}
\cF^{(d)}_{\{\pi_{f^d},\pi_{f^{{\phantom |\!\!}}},\pi_{f^d}\}}(q,u) \ =\ \frac1{(d!)^2}\,\varphi_{-2,1}(\tau,z)^{d-1} {1\over \Delta(\tau)}\,,\quad d\geq1\,,
\ee
with 
\be\label{pm21}
\varphi_{-2,1}(\tau,z)\ =\ {\vartheta_1(\tau,z)^2\over \eta(\tau)^6}\ =\ u^2+\cO(u^4)\,,
\ee
where we substituted  $q=e^{2\pi i\tau}$, $u\equiv-{2\pi i} z$.
Before we will discuss some physical implications,
let us first make a few brief remarks:
\bi

\item For $d=0$, (\ref{FOb}) turns formally into the well-known KKV formula 
\cite{Katz:1999xq}. It typically arises as denominator of elliptic genera of $\sym^d(K3)$,
where it reflects the orbital contributions of BPS states, and
its poles play an important role for black hole physics \cite{Katz:1999xq,Haghighat:2015ega}.
However, for us where $d\geq1$ the physics is different: there aren't any poles, 
and the blow-up mode $u$ of the symmetric product exists only for $d\geq2$.

Indeed, for $d=1$ which corresponds to a single $K3$, 
all $u$-dependence disappears from (\ref{FOb}). 
It turns into the Yau-Zaslow formula 
$\sum_{h\geq0}\langle\, 1 \ \rangle^{K3}_{0,0;\beta_h}\, q^{h-1}=\Delta(q)^{-1}$
which counts genus zero curves in the $K3$ \cite{Yau:1995mv}.
This reflects that we count here only those higher genus curves that are induced
by the perturbation in the blow-up mode $u$;
in a sense the extra inverse power of $\varphi_{-2,1}(\tau,z)$ in (\ref{FOb}) cancels out the
unwanted higher genus curves in $K3$. 

That the $u$-dependence of the $(2d+1)$-point function (\ref{Fqu}) 
drops for $d=1$ is also consistent with non-renormalization theorems \cite{deBoer:2008ss,Baggio:2012rr},
which state that quadratic and cubic couplings are not renormalized.

\item
The function in (\ref{pm21}) is a standard generator of the ring of Jacobi forms,  $\varphi_{w,m}: \IH\times \IC\rightarrow\IC$,
of modular weight $w=-2$ and index $m=1$. 
As such it shares the defining feature of Jacobi forms, namely their modular and
double periodic transformation properties,
\bea\label{Jacmodular}
\varphi_{w,m}  \left(\frac{a \tau + b}{c \tau +d}, \frac{z}{c \tau +d} \right) &=& (c \tau+d)^w e^{2\pi  i  \frac{m c}{c\tau +d}  z^2}    \varphi_{w,m}(\tau,z)\ \ \,{\rm for}    
 \left(\begin{matrix} 
      a & b \\
      c & d \\
   \end{matrix}\right) \in SL(2,\IZ),\ \ 
\\
\varphi_{w,m}\left( \tau , z + \lambda \tau + \mu \right) &=& e^{-2 \pi i   m (\lambda^2 \tau  + 2  \lambda  z )  }   \varphi_{w, m} (\tau,  z)\,,
\quad \lambda, \mu \in \mathbb Z \,,\label{periodicity}
\eea
which has have important physical implications as will see further below. 

\item
Given that $\varphi_{-2,1}(\tau,z)$ is essentially a theta function, we can make use of its various canonical expansions, 
thereby using different parametrizations in terms of $y,u$ or $z$, etc, whichever fits best.
In  particular we can write
\be 
\varphi_{-2,1}(q,u)\ =\ u^2 \,{\rm exp}\,\left({\sum_{g\geq1}
\frac{B_{2g} }{g (2g)!}E_{2g}(q) u^{2g}}\right).
\ee
Here $E_{2g} (q)=1+\cO(q)$ are the familiar Eisenstein series, which are (quasi-)modular forms of weight $w=2g$,
and $B_{2g}$ are the Bernoulli numbers.
In the limit of large $K3$ volume, $q=0$, this simplifies to the ubiquitous sine function
\bea
\varphi_{-2,1}(0,u)&=& \cS(u)^2\ =\ u^2+\cO(u^4), \ \ \ {\rm where}\\
\cS(u) &:=&  2\, \sinh\left(\frac u2\right)\ =\  -i(y^{1/2}+y^{-1/2})\,,\nn
\eea
whose enumerative significance in GW/Hurwitz theory is well known. In terms of it,
we can express $\varphi_{-2,1}$ and thus the generating function in Gopakumar-Vafa form \cite{Gopakumar:1998ii,Gopakumar:1998jq}:
\be\label{FGV}
\cF^{(d)}_{\{\pi_{f^d},\pi_{f^{{\phantom |\!\!}}},\pi_{f^d}\}}(q,u) \ =\ \frac1{(d!)^2} \sum_{g,h\geq0} \GV_{g,\beta_h,d}\, q^{h-1}\,\cS(u)^{2g-2+2d}\,,
\ee
where $\GV_{g,\beta_h,d}$ are integer invariants that count BPS states.
This way of writing makes the advertised
$y\rightarrow 1/y$ symmetry manifest which is not visible in a Laurent expansion around $y=0$ (resp.~$\infty$).

\item
There are the following canonical expansion points, namely the perturbative genus expansion
around the orbifold point $u=0$ resp.~$y=-1$, potential
strong-coupling expansions around the singularities at $y=\{0,\infty\}$, plus possible scaling limits involving
concurrent tuning of the $K3$ K\"ahler modulus. 
We will discuss the genus and strong coupling expansions in more detail in the next two sections.

\item
As a 
side remark, note that $2d+1$-point functions of K\"ahler moduli being encoded in the 
quantum cohomology of $\sym^d(K3)$ fits
structurally together with findings made in refs.~\cite{Lerche:1998gz,Lerche:1999hg}. The physical context there was
very  different, 
namely it was about computing certain BPS-saturated, quartic gauge and gravitational couplings in 
$D=8$ dimensions, which by supersymmetry relate to quartic derivative scalar couplings (or to quintic ones by taking an
extra derivative, which fits better to the present discussion). 
The idea was to compute the couplings in the
heterotic string duality frame at one-loop order, and then to translate the result to F-theory to see  what
geometrical structure of the elliptic $K3$ this would correspond to. 

It was found that the couplings were governed
by Picard-Fuchs equations pertaining to $\sym^2(K3)$, with extra source terms that looked like as if coming
from a five-fold with $\sym^2(K3)$ as fiber. Attempts to find such a five-fold were not successful, but
with hindsight we would say today that what was computed were in fact {\it reduced} invariants
related to the twistor family of $\sym^2(K3)$; this notion was not known at the time.

Moreover, based on detailed computations \cite{Lerche:1999hg} it was conjectured that this pattern is 
more general, in
that formally in $D=4d$  dimensions, certain protected $2d+1$-point amplitudes of gauge/moduli fields are geometrically
characterized by $\sym^d(K3)$. This mirrors the structure that we find here.
\ei

\subsection{Perturbation theory near the orbifold point} \label{sec_perturbation}

Our example exhibits the highly degenerate nature of the theory at the orbifold point, 
$u=0$: all ramifications
are trivial, and the ``world-sheet'' is maximally disconnected
(cf., Figure~\ref{fig:unramified2}).  It consists of $d$ copies of  $\IP^1$'s, 
which we may interpret as decoupled world-sheets of a collection of  tensionless strings
that ought to arise in the undeformed theory. Each comes
with two insertions
of the K\"ahler form $f$ of the elliptic fiber of $K3$, except one that comes with an extra insertion for stability.
In total this gives a completely factorized, \mbox{$(2d+1)$}-point function of
K\"ahler moduli, whose connected component vanishes identically.

This is reflected by the overall factor $u^{2d-2}$ of the generating function (\ref{Fqu}): one needs to take
$2\cdot(d-1)$ $u$-derivatives in order to arrive at a connected genus $g=0$ invariant. 
This amounts to inserting in total
$2d-2$ twist fields such as to glue the $d$ sheets together along $d-1$ branch cuts. One may view this as a binding process that merges $d$ single  tensionless strings wrapped once along 
$\IP^1=\partial {\rm AdS}_3$
into a single, $d$-times wrapped long string, with $2d+1$ insertions of massless
K\"ahler moduli.  This realizes a picture that had been advocated in \cite{Dijkgraaf:1998xr}.
The details how this works are a bit subtle and will be discussed below.
Going the other way, ie., approaching $u=0$ from non-zero $u$, is nothing but
the fragmentation process described in \cite{Seiberg:1999xz} according to which branes can separate without
cost in energy when the background fields are switched off.

The combinatorics of twist fields at arbitrary orders in conformal perturbation theory is a priori complicated, but here all is encoded  in the $u$-expansion of the Jacobi form $\varphi_{-2,1}(q,u)$.
Since for the expansion near $u=0$ the K\"ahler modulus is not important, we set $q=0$ for now;
in the next section we will discuss some extra features that arise when $q\not=0$.

Dropping $q$ one expects the generating function (\ref{Fqu}) to be governed by Hurwitz theory (for a review,
 see eg., \cite{Lando2011HurwitzNO}),
as we count branched coverings of $\IP^1$ after all.
At first sight, the relevant Hurwitz numbers ought to be the so-called simple Hurwitz numbers $H_{g,d}^{(\mu)}$
that count covering maps of degree $d$ with given 
ramification $\mu=\{\mu_i\}$ over one fixed point,
plus $N(g,d,\mu)= 2g-2+d+\ell(\mu)$  simple branch points.
As indicated in Appendix~\ref{app_topgr}, Hurwitz numbers can be written as correlators in topological gravity on $\IP^1$ \cite{Pandharipande:1999zry}, and concretely for simple Hurwitz 
numbers we have:
\be\label{Hddtau}
H_{g,d}^{(\mu)}\ =\ \big\langle\,\mu\,|\,{\tau_1(\omega)}^{2g-2+2d}\,|\,1^d\,\big\rangle^{\IP^1/\{0,\infty\}}_{g,d}\,,
\ee
which can be summed up to a generating function:
\be\label{Huruv} 
H_d^{(\mu)}(u)\ =\ u^{2d-2}\sum_{g\geq0} 
\frac{u^{2g} }{(2g+2d-2)!}H_{g,d}^{(\mu)}\,.
\ee 
However, for our situation with
trivial ramifications so that $(\mu)=(1^d)$, this does not match the  (\mbox{$q$-independent} part of the) generating function (\ref{Fqu}) for $K3\times\IP^1$ which reads:
\bea\label{Fqunull}
\cF^{(d)}(0,u)& :\equiv&
q\,(d!)^2 \,\cF^{(d)}_{\{\pi_{f^d},\pi_{f^{{\phantom |\!\!}}},\pi_{f^d}\}}(q,u) \big\vert_{q=0}
\nn\\
&=& \label{FitoS}
   \cS(u)^{2d-2}\,.
\eea
An exception is for $d=2$ where:
\be\label{S2H2}
\cS(u)^2\ =\ 4\, H_2^{(1^2)}(u)\,,
\ee
and trivially for $d=1$. Thus the most direct connection to simple Hurwitz numbers can be stated as
$\cF^{(d)}(0,u)=2^{2d-2}(H_2^{(1^2)}(u))^{d-1}$, which reflects the factorized geometry.

\begin{figure}[t!]
\centering
\includegraphics[width=15cm]{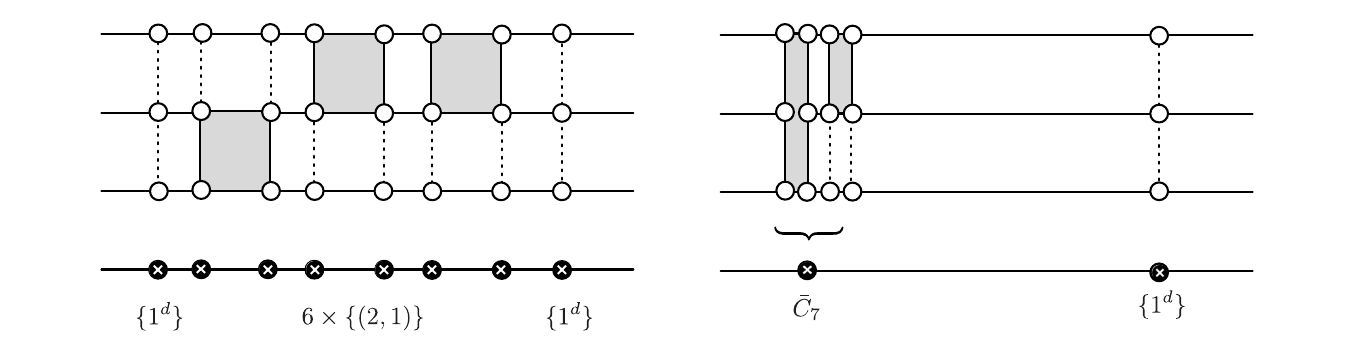}
\caption{
Examples of connected genus $g=1$,  degree $d=3$ branched covers of $\IP^1$.
\\
Left: 
Genus one curve arising from connecting the sheets by $2g-2+2d=6$ simple branch points.
The number of such coverings is given by the simple Hurwitz number $H^{(1^3)}_{1,3}=\langle\tau_1(\omega)^6\rangle^{\IP^1}_{1,3}=40$.
This is however not the geometry we encounter in the perturbation expansion at order $u^6$,
rather it corresponds to what we call a bare curve, $\Sigma^{(0)}$, at $u=0$.
\\
Right:  Crude sketch of the perturbed geometry at order $u^6$. 
It corresponds to an insertion of a completed cycle with associated Hurwitz number
$\overline{H}^{(1^3)}_{1,3}(\overline{C}_7)=\langle{\tau_6}(\omega)\rangle^{\IP^1}_{1,3} =\frac5{864}$. 
By misuse of notation, we drew it in terms of slightly separated simple ramification points, but actually they all lie on
top of each other so that the genus $g=1$ curve contracts to a single branch point over the base. 
}
\label{fig:combo}
\end{figure}

This may seem odd as it is well established \cite{Pakman:2009ab} that certain symmetric
orbifold correlators are given by simple Hurwitz numbers $H^{(\mu)}_{g,d}$ 
(at $g=0$ for extremal correlators).
How does this tie together with our discussion?

In fact, simple Hurwitz numbers (\ref{Hddtau})
do appear when counting covering maps for the 
bare curves $\Sigma^{(0)}$ at $u=0$ \cite{Pakman:2009ab}.
For these, all insertions are tied to fixed branch points $z_j$ of $\IP^1$ -- see the left hand side of Figure~\ref{fig:combo}.
 However, the perturbation theory in $u$ we consider corresponds to
twist field insertions, $\int \hG^-{\bar\hG}{}^-\sigma_2$, that are integrated over. 
Thus degenerate situations occur where twist fields collide with each other or with the relative insertions $\gamma^{(\pi^{(j)})}(z_j)$.  It is known \cite{Dijkgraaf:1990qw} that in topological gravity
 integrated descendant insertions contribute 
via contact terms, and no extra markings at fixed locations are introduced. Thus it is no
surprise that the simple Hurwitz numbers (\ref{Huruv}) do not appear in the perturbation expansion.

What is then the enumerative significance of the genus expansion of $\cF^{(d)}(0,u)= \cS(u)^{2d-2}$?
It is revealed by the appearance of the sine function: for Gromov-Witten theory of $\IP^1$ 
it is known \cite{OkouPandha0204305} to arise as generating function for Hurwitz numbers
$H_{d}(\overline{C}_{k+1})$ of ``completed'' cycles, $\overline{C}_{k+1}$. 
Such completed Hurwitz numbers, also called $k+1$-spin Hurwitz numbers \cite{Goul_math/0309440,duninbarkowski2020loop},
are generalizations of the simple Hurwitz numbers where simple ramifications are replaced
by ramifications with completed cycles. 

As briefly mentioned in Appendix~\ref{app_topgr}, these cycles are labeled by certain
distinguished linear combinations of partitions, namely precisely those
that are equivalent to insertions of the gravitational descendants:
\mbox{ $\tau_k\equiv {\tau_1}^k\sim\frac1{k!}\overline{C}_{k+1}$.}
The coefficients of the linear combinations are given in terms
of the sine function. Specifically, denoting by $(\mu)$ the conjugacy class of a partition $\mu=\{\mu_i\}$ of $d$, 
the completed cycles are given by linear combinations~\cite{OkouPandha0204305} 
\be
\overline{C}_{k+1}\ =\ \sum_\mu \rho_{k+1,\mu}\,(\mu)\,,
\ee
where
\be\label{rhodef}
\rho_{k+1,\mu}\ =\ \frac{k!}{d!}\prod \mu_i\left(  \cS^{d-1}(u) \prod\cS(\mu_i u)\right)\Big|_{u^{k+1}}\,,
\ee
is the source of the sine function.
The completed Hurwitz numbers are then given by
\be
\overline{H}^{(\mu)}_{g,d}(\overline{C}_{2g-1+d+\ell(\mu)})\ =\ \big\langle\,\mu\,|\,{\tau_{2g-2+d+\ell(\mu)}(\omega)}\,|\,1^d\,\big\rangle^{\IP^1/\{0,\infty\}}_{g,d}\,,
\ee
and in our situation with trivial ramification, $(\mu)=(1^d)$, one obtains
\cite{Pandharipande:1999zry}
\be\label{P1only}
u^{2d-1}\sum_{g\geq0} u^{2g}\,\overline{H}^{(1^d)}_{g,d}(\overline{C}_{2g-1+2d})\ =\
\frac1{(d!)^2}\,\cS(u)^{2d-1}\,.
\ee

To give a geometrical interpretation, recall that the sine function prominently appears in topological strings as summing
degenerate (pinched) curves as per the GV formula  (\ref{FGV}}. This is what is reflected here, as we can think\footnote{I thank Vincent Bouchard for explanations.} of a completed cycle as a degenerate configuration where a genus $g$ curve carrying $d$ marked points contracts via the covering map to a single point in the base $\IP^1$; see the right part of Figure~\ref{fig:combo} for a crude visualization. We can view
this as the result of the transformation (\ref{downtoone}). Recall, however, that
the representation of the GW invariants is not unique and that there
are in general several different but mathematically equivalent representations. 
For example one can distribute the genus $g$ over curves contracted over several branch points.

Note that the foregoing formulas refer to topological gravity with target $\IP^1$ rather than $K3\times\IP^1$. 
The generating functions are closely related in that there is an extra power of $\cS(u)^{-1}$ in 
 (\ref{FitoS}) as compared to (\ref{P1only}).   
 To understand this, recall that  the GW invariants we consider momentarily
  count maps into curve classes
 $({\rm pt},d)$ of $K3\times\IP^1$, 
where $d$ is the degree of a map to $\IP^1$. These include also constant ($d=0$)  maps from collapsed curve components into $K3$. For example, 
such a map from a genus $g=g_1+g_2$ curve can involve a degree $d$ map from the genus $g_1$ component to $\IP^1$ and 
a degree zero map from a collapsed genus $g_2$ component to a point on $K3$.
Thus in general there is an extra contribution from degree zero maps to $K3$, which in the present context
is given by an integral over an extra, one-pointed component
$\overline{ \cM}_{g_2,1}$ of the compactified moduli space. It can be written as a Hodge integral \cite{Faber:1998gsw}
\be\label{degreezerocontr}
u^{-1}\left[1+ \sum_{g_2\geq1} u^{2g_2}\int_{\overline{ \cM}_{g_2,1}}\lambda_g \psi^{2g_2-2}\right]\ =\ \cS(u)^{-1}\,,
\ee
where $\lambda_i=c_i (\Lambda^{\!\vee}) \in H^{2i}(\overline{\cM}_{g,n})$ are Chern classes
of the dual Hodge bundle, and $\psi$ is the representative of $\tau_1$ under the integral as mentioned before.
Together with the contribution (\ref{P1only}) from $\IP^1$, this reproduces the generating function
(\ref{Fqunull}) for $K3\times\IP^1$ at $q=0$.

Wrapping up: our purpose in this section was to elucidate 
the genus expansion in the blow-up modulus $u$ near $q=0$. 
We argued that a natural interpretation is in terms of degenerate, 
contracted genus $g$ curves that are attached to the bare curves $\Sigma^{(0)}$ at their branch points.
This supports the viewpoint of contact terms: the 
genus corrections are integrated out and are subsumed by ``renormalized" insertions, given by completed cycles. This
is reminiscent of flat operator representatives in topological Landau-Ginzburg
theory which encode renormalization due to integrating out massive states.
This viewpoint also offers an appealing physical picture  where the 
string world-sheet touches the boundary of AdS$_3$ only at a few fixed branch points 
and not at extra marked points whose number would depend on the order of perturbation theory.

\subsection{Large-distance limit} \label{sec_largeD}

We now turn to a potentially more interesting limit, namely $y= -e^{-u}= -e^{2\pi i z}\rightarrow0$  
(which is equivalent to $y\rightarrow\infty$ due to the $y\leftrightarrow1/y$ symmetry),
while restoring the dependence of the fibral modulus $\tau$ of the $K3$.
From the viewpoint of a flat coordinate on the CFT moduli space this corresponds to a large
distance limit, which by the  lore of the Swampland Distance Conjecture \cite{Ooguri_2007}
ought to correspond to a weak coupling limit in some proper duality frame.

However, the situation is different as compared to the standard large distance limits in
K\"ahler moduli space. This is because $u$ as elliptic parameter of a Jacobi form  lives on the
Jacobian torus and thus is essentially a double periodic variable. More precisely, our generating
function, which is a Jacobi form with non-zero index, transforms with an 
``anomalous'' prefactor (\ref{periodicity}), that is,
\be\label{doubleP}
\cF^{(d)}_{\{\pi_{f^d},\pi_{f^d}\}}(\tau,z+\lambda\tau+\mu) \ \propto\ 
\varphi_{-2,1}(\tau,z+\lambda\tau+\mu)^{d-1} \ =\ 
q^{-\lambda^2(d-1)}(-y)^{-2\lambda(d-1)}\varphi_{-2,1}(\tau,z)^{d-1}\,,
\ee
with $\lambda,\mu\in\IZ$. 
The periodicity in $\mu$ corresponds to the familiar shift of the theta-angle, while the
the quasi-periodicity in $\lambda$ is less obvious. Such a shift can always be interpreted in
terms of spectral flow\footnote{The period of this spectral flow 
is given by the $K3$ fibral K\"ahler parameter 
$\tau$ so it appears different to the one familiar from the literature \cite{Maldacena:2000hw}.} related to a $U(1)$ symmetry whose associated field strength,
or fugacity parameter, is given by~$z$.    Altogether there seems a whole lattice
 $\IC/(\IZ\tau \times\IZ)$ of zeros that allow for weakly coupled genus expansions
 in terms of suitably spectrally flowed degrees of freedom that become light there.

However, due to the prefactor in the transformation law (\ref{doubleP}), the situation is not
really periodic. That is,
if we increase $z=iu/2\pi$ arbitrarily far in the $\tau$-direction, we can always map it back to the
fundamental domain
\be\label{fundam}
F_z \ =\ \big\{\,z\,|\, 0\leq{\Im z}\ <\  {\Im \tau}\big\}\,,
\ee
albeit at the expense of a divergent prefactor. This just reflects the lack of convergence outside
the fundamental domain. This can also be seen by considering the limit of (\ref{FGV}) as~$y$ becomes small,
after factoring out the leading singularity $y^{1-d}$:
\be
q\,\cF^{(d)}_{\{\pi_{f^d},\pi_{f^{{\phantom |\!\!}}},\pi_{f^d}\}}(q,y)\ \propto\ y^{1-d} \sum (-1)^g\GV_{g,\beta_h,d}\,  \,     y^{-g}\,q^h\,.
\ee
Then one may worry about the convergence of the second factor. 
It is a property of $\varphi_{-2,1}$ that its Gopakumar-Vafa expansion coefficients obey
\be
GV_{g,\beta_h,2}\ =\ 0\,\ \ \ {\rm for}\ g>h\,,
\ee
so that the leading terms in  $q/y=-e^{2\pi i(\tau-z)}$  take the form:      
\be
q\,\cF^{(d)}_{\{\pi_{f^d},\pi_{f^{{\phantom |\!\!}}},\pi_{f^d}\}}(q,y)\ \propto\ y^{1-d}\left[ \sum (-1)^g\GV_{g,\beta_g,d}\, \,\Big(\frac qy\Big)^g\!+\cO\Big(\frac {q^2}y\Big) \right]\ \propto\ y^{1-d}\left[\left(1+\frac qy\right)^{2d-2}\!\!\!+\cO\Big(\frac {q^2}y\Big) \right]
\,.
\ee
The second factor indeed converges over the fundamental domain, $F_z$. 

The upshot is that even for large fiber volume\footnote{Note that (for the F1NS5 system) the total volume of the $K3$ 
in the near horizon limit is linked to the ten dimensional string coupling
via ${g_{10}}^2=d^{-1}{\rm Vol}(K3)/{\alpha'}^2$ \cite{Giveon:1998ns,Berkovits:1999im,Larsen:1999uk, OhlssonSax:2018hgc}.
 We can choose the volume as a free parameter at the expense of
a volume dependent ten dimensional string coupling. Thus with increasing volume the
theory runs into a non-perturbative regime
at $g_{10}=1$, beyond which one needs to switch to the S-dual D1D5 frame \cite{Martinec:2022okx}. 
If we want to stay in the current framework, this puts a bound on
the volume, Vol$(K3)/{\alpha'}^2<d$, and thus morally speaking on ${\rm Im}\tau$. 
Strictly speaking 
Im$\tau$ gives the volume of the elliptic fiber of the $K3$ only, and the total volume will depend also on
the volume of the base of the fibration; this subtlety is not important for our purposes.
We just need to keep this bound in mind when talking about large volume limits.
}
of the $K3$, Im$\tau\rightarrow \infty$, one can take Im$z$ large only as far as one stays within the domain of convergence $F_z$. The lack of convergence outside $F_z$ signals
that new degrees of freedom should come into play that were not taken care of in the
present framework. This calls for a non-perturbative completion of the theory,
quite likely in the form of a matrix model \cite{Gaberdiel:2020ycd}.  

That a string coupling enters as the elliptic argument of a Jacobi form and
so exhibits \mbox{(quasi-)}periodicity, is a common feature in topological strings \cite{Katz:1999xq,Huang:2015sta,Haghighat:2015ega},
one difference being  that we deal here with zeros  rather than with poles at the
integral lattice points. There has been recent progress in 
non-perturbative completions (see eg., \cite{Marino:2015nla,Alim:2021mhp,Gu:2023mgf}) 
of topological strings in terms of resurgence, and it would be natural to expect that analogous methods would
work here too.

As for  Swampland physics, it is unfortunate that the regime of large Im$z$  lies outside the domain of 
validity of the framework we developed here, and the answer to important questions, 
like about the appearance of higher spin states and
scale separation, seems to be buried in the details of the purported non-perturbative completion.
On the other hand, if we stick to the regime $F_z$  (\ref{fundam}) where convergence is
under control, Im$z$ cannot grow larger than Im$\tau$ and this translates,
via (\ref{Rads}), to the statement that $R_{{\rm AdS}}$ is bounded from above
by the size (of the elliptic fiber) of the $K3$. 
In other words, scale separation, in the sense we just explained, does not occur within the present framework.

 \section{Discussion} \label{sec_summary}

 In this paper we argued how certain results in algebraic geometry 
 \cite{Ober_1406.1139,Oberdieck:2014aga,Ober_1605.05238,Oberdieck:2022khj}
could provide insights in topological aspects of the AdS/CFT correspondence for $AdS_3\times S^3\times K3$ for \mbox{$Q_5^{NS}=1$}. Our emphasis was 
 on the moduli space and the perturbation expansion in the blow-up
 deformation parameter, $u$, of the symmetric orbifold $\sym^d(K3)$ on the CFT side.
 We showed that it corresponds to a 't~Hooft-like expansion around
 a given background geometry, defined in terms of a bare curve, $\Sigma^{(0)}$, which 
  is specified by a  given set of (in general cohomology weighted) ramification
 conditions, $\{\pi_j\}$. The effect of this expansion is to add connections between a priori
 disconnected sheets and/or adding more and more handles to $\Sigma^{(0)}$.
Near the orbifold point, $u=0$,  it takes the form of a perturbative genus expansion for the 
tensionless strings that dominate this regime.
 On the other hand, in the large distance limit $u\rightarrow\infty$, the framework breaks 
 down and ought to be supplemented by a suitable non-perturbative completion.
We argued that within the domain of convergence, scale separation does not occur.

However, as pointed out at various stages of this work, there is an important issue when 
attempting to connect, as per Figure~\ref{fig_correspondences}, the mathematical Gromov-Witten/Hilbert Correspondence to the physical AdS/CFT Correspondence.
The root of the problem lies in the fact that GW theory deals with holomorphic geometry while a hyperk\"ahler manifold such as $\hilb^d(K3)$ has a natural harmonic rather than holomorphic structure. 
In order to get non-zero invariants in the first place, one
considers in the mathematical literature reduced Gromov-Witten invariants  \cite{Bryan1997TheEG, Maul_0705.1653,Maulik:2010jw,Klemm_0807.2477,Pandharipande:2014ooa},
which per design count holomorphic maps in hyperk\"ahler geometry.
While in physical language these ought to count certain BPS invariants,  it is
a priori not clear what actual physical correlation functions on the AdS side they would correspond to.

We have argued how to interpret these reduced 
invariants in the language of
topological $N=4$ strings, as defined in \cite{Berkovits:1994vy,Berkovits:1999im}.
At first sight, this seems to fit to the recent works on the AdS side (e.g., \cite{Dei:2020zui,Bhat:2021dez,Gaberdiel:2022als,Knighton:2022ipy,McStay:2023thk}) which indeed
point to topological $N=4$ strings; their correlators boil down to counting holomorphic covering maps of the boundary,
and thus rest on an embedding of holomorphic into hyperk\"ahler geometry as well.
If this interpretation bears out, 
reduced invariants for $m\geq4$ insertions can be written as physical correlators of the form $\langle\phi\phi\phi\int\bar \phi \int \hG^{-}\phi\dots\! \int \hG^{-}\phi\rangle$, for details see  around eq.~(\ref{noholcorr}).
%

Evidently there are loopholes in this line of thought. In particular,
$N=4$ strings are not unique in the sense that
there is a choice of which combinations of $\hG{}^\pm(\zeta)$ versus $\htG{}^\pm(\zeta)$
are inserted in correlation functions. This can be translated to the language of
picture changing, and the prescription of \cite{Berkovits:1994vy,Berkovits:1999im} is to sum over all pictures,
or possible combinations of such insertions. Now, in the
concrete hybrid formulation of $N=4$ strings on AdS$_3\times S^3\times M_4$
alluded to above, there are extra constraints on correlators due to ghost number conservation,
and these  restrict which combinations of insertions of $\hG{}^\pm(\zeta)$, $\htG{}^\pm(\zeta)$  can yield non-vanishing contributions. In other words, the issue is whether the localization to the boundary
$\IP^1=\partial$AdS$_3$ properly aligns the complex structure parametrized by $\zeta_i$
such that it leads to non-trivial reduced invariants.

While in ref.~\cite{Fiset:2022erp} marginal deformations of genus zero correlators with an insertion of $(\tG^+)^{-1}$
have been considered recently, it is
not clear at this point whether our specific interpretation of reduced
invariants is compatible with the concrete realization of $N=4$ strings
that is relevant for AdS$_3\times S^3$. There might be a subtlety in defining the theory,
analogous to the choice in topological gravity whether to consider the ``small'' or the ``large'' 
Hilbert (or phase) space, depending on the definition of physical states in terms of equivariant cohomology \cite{Verlinde:1990ku,Dijkgraaf:1990qw};
the latter choice includes gravitational descendants, while the former does not.
This analogy is suggested by the fact that the genus expansion we consider here is implemented by insertions of the twist field~$\sigma_2$, which via the Gromov-Witten/Hurwitz Correspondence (cf., Appendix~\ref{app_topgr}) is equivalent to insertions of the gravitational descendant~$\tau_1$.
In a nutshell, essentially this boils down to the question whether the topological
$N=4$ strings on AdS$_3\times S^3\times K3$
really localize to topological strings on $K3\times \IP^1$ that admit such a ``large'' Hilbert space, or not.

Clearly,  in order to put the suggested connection between the GW/Hilbert and the AdS/CFT Correspondence on a firmer ground,  a more detailed analysis of correlation functions and their marginal deformations will be necessary.
These matters also have  a bearing
on non-renormalization theorems such as discussed in \cite{deBoer:2008ss,Baggio:2012rr}.
We leave this to future work.

 \vfil
 
\subsection*{Acknowledgments} 
I thank Murad Alim,  Vincent Bouchard, Jim Bryan, Lorenz Eberhardt, 
Alba Grassi, Bob Knighton, Shota Komatsu, Marcos Marino, Georg Oberdieck, Kyriakos Papadodimas and Timo Weigand for discussions and/or help, and especially Matthias Gaberdiel also for hospitality at ETH and Cumrun Vafa for
an important remark.
Moreover I am grateful to the TH Department at CERN for support and hospitality.

\appendix
\section{GW/Hurwitz Correspondence and topological gravity} \label{app_topgr}

Let us briefly elucidate some of the underlying mathematical structure
on the ``Gromov-Witten'', right-hand side of (\ref{GWhilb}), which at least in principle may facilitate explicit computations.

Branched coverings of $\IP^1$ are well-known to be the subject of Hurwitz theory, 
and what is relevant here is its correspondence \cite{OkouPandha0204305}
 with Gromov-Witten theory of~$\IP^1$.
The underlying mathematical logic is an equivalence of ramification profiles
and intersection theory of Riemann surfaces, which in the language of topological gravity 
\cite{Witten:1990hr,Verlinde:1990ku} is formulated in
terms of gravitational descendants,  ~$\tau_k\equiv{\tau_1}^k$.
To leading order these correspond to the twist fields: $\tau_k\sim\frac1{k!} \,\sigma_{k+1}$, but in general there are
extra correction terms such that $\tau_k$ correspond to a special basis of
``completed cycles'' \cite{OkouPandha0204305}.
These reflect degenerations of Hurwitz coverings, which amount to contact terms
in physics language; see the end of Section~\ref{sec_perturbation} for comments
in the context of an example. There are no corrections for~$\tau_1=\sigma_2$,
which is associated with the blow-up mode of the orbifold singularity.

More specifically, what is relevant here is an extension of Hurwitz theory  to cohomology weighted partitions
that was dubbed ``relative/descendent correspondence'' in \cite{Pand_1011.4054}.
It maps  between the Nakajima Fock space basis (labeled by cohomology weighted partitions $\pi=[(\mu_i,\alpha_i)]$)
and descendant insertions as follows:
\bea
 \label{gammatau}
\gamma^{(\pi)} &\rightarrow& \gamma^{(\pi)}[\tau]\ =\ \sum_{\pi'} (M^{-1})_{\pi\pi'}\, \tau^{(\pi')}\,,
\\
{\rm where} \qquad\qquad
\tau^{(\pi)}& :=& \prod_{i=1}^{\ell(\pi)} {\tau}_{\pi_i-1}(\alpha_i\,\omega)\,,
 \qquad\qquad \qquad\qquad
\eea
with  $\omega\in H^2(\IP^1)$.
The linear basis transformation is given by a matrix of relative one-point overlap invariants of the schematic form $M^{\pi'\pi}=\mathfrak{z}(\mu)\langle\gamma^{(\pi)}|\tau^{(\pi')}\, \rangle^{K3\times \IP^1\!/\infty}$\!, which can be shown to be invertible; 
details are spelled out in refs.~\cite{OkouPandha0204305,Pand_1011.4054,Oberdieck:2021ftp,Oberdieck:2021beb}.
This allows to express invariants with ramification conditions $\mu^{(j)}$ at locations $z_j$ in terms of correlators of topological gravity and vice versa. Concretely one can represent the invariants (\ref{GWK3P1}) as linear combinations of
\be\label{finaltop}
 \sum_{l_j\geq0\atop \sum \l_j= N(d,g,\mu)}\big\langle \,
\prod_{j=1}^n \tau_{l_j} \big( \prod_{i=1}^{\ell(\mu^{(j)})}{\tau}_{\mu^{(j)}_i-1}(\alpha^{(j)}_i\,\omega)\big)\,
\big\rangle_{g;(\beta,d)}^{K3\times\IP^1/\{z_1,\dots z_n\}} \,,
\ee
where $N(g,d,\pi)=2(g-g_0)$ is the genus deficit as defined in (\ref{Ndef}).

In fact, by repeated application of (\ref{gammatau}) one can map all relative insertions to just one insertion (\cite{Oberdieck:2021ftp}, Sect. 5.2), schematically:
\be\label{downtoone}
\big\langle\, \prod_{j=1}^n\gamma^{(\pi^{(j)})}\,\big\rangle^{K3\times\IP^1/\{z_1,\dots,z_n\}} 
\ \ \ \longrightarrow\ \ \ 
\big\langle \,\gamma^{\pi'}\,
\prod_{i=1}^{\ell(\tilde\pi)} {\tau}_{\tilde\pi_i-1}(\alpha_i\,\omega)\,
\big\rangle^{K3\times\IP^1/\infty} \,.
\ee
This representation is highly degenerate as the whole genus $g$ curve contracts over one point of $\IP^1$;
see again the end of Section~\ref{sec_perturbation} for an example.

Thus, in general there is a large degree of 
ambiguity in representing the reduced GW invariants for $K3\times \IP^1$.
Specifically, it can be shown  by localization and degeneration arguments~\cite{Ober_1605.05238} 
that the full Gromov-Witten theory of $K3\times\IP^1$ can be expressed in terms
of ELSV-type \cite{Ekedahl_1999,Ekedahl_2001}, linear Hodge integrals for $K3$ only. Conversely,
Hodge integrals are completely determined in terms of descendant invariants \cite{Faber:1998gsw}.


\bibliography{papers}
\bibliographystyle{utphys.bst}

\end{document}